\documentclass[12pt]{article}

\usepackage{amssymb,amsthm,amsfonts,psfig,epsfig, multicol,mathrsfs,axodraw}
 
\def\BrightRed  {}
\def\Black{}
\def\Green{} 
\def\Blue {}
\def\gappeq{\mathrel{\rlap {\raise.5ex\hbox{$>$}} {\lower.5ex\hbox{$\sim$}}}}
\def\lappeq{\mathrel{\rlap{\raise.5ex\hbox{$<$}} {\lower.5ex\hbox{$\sim$}}}}
\def\beq{\begin{equation}} \def\eeq{\end{equation}} 
\def\bea{\begin{eqnarray}} \def\eea{\end{eqnarray}}
\def\bq{\begin{quote}} \def\eq{\end{quote}}
\def\bc{\begin{center}} \def\ec{\end{center}}

\def\nn{\nonumber}

\begin{document} 


\pagestyle{empty}
\def\thefootnote{\fnsymbol{footnote}}
\begin{flushright} ROMA1-TH/1427-06  
\end{flushright}
\vskip 2cm

\bc {\Large \bf  
Real and Imaginary Elements \\ \vskip .4cm  
of Fermion Mass Matrices}  \ec
\vspace*{2cm} 

\centerline{ \large \bf I. Masina$^a$ and C.A. Savoy$^b$ } 
\vskip 1cm

\bc { \em a) Centro Studi e Ricerche "E. Fermi", Via Panisperna 89/A, Rome, Italy \\ \vskip .1cm  
and  INFN, Sezione di Roma, P.le A. Moro 2, Rome, Italy} \\ \vskip.2cm
{ \em b) Service de Physique Th\'eorique, CEA-Saclay, Gif-sur-Yvette, France } \ec

\vskip 2cm

\centerline{\bf Abstract} 
\vskip.2cm
Prompted by the recent better determination of the angles of the unitarity triangle,
we re-appraise the problem of finding simple fermion mass textures, 
possibly linked to some symmetry principle and compatible with grand unification.
In particular, the indication that the angle $\alpha$ is close to rectangle
turns out to be the crucial ingredient leading us to single out 
fermion mass textures whose elements are either real or purely imaginary. 
In terms of the five parameters ascribed to the quark sector,
these textures reproduce the eight experimental data on quark mass ratios and mixings 
within $1\sigma$. 
When embedded in an $SU(5)$ framework, these textures suggest a common origin
for quark and lepton CP violations, also linked to the spontaneous breaking of the gauge group.


\newpage
\setcounter{page}{1}
\pagestyle{plain}
\def\thefootnote{\arabic{footnote}}
\setcounter{footnote}{0}


{\it "Perche', secondo l'opinion mia, 
a chi vuole una cosa ritrovare, 
bisogna adoperar la fantasia, 
e giocar d'invenzione, e 'ndovinare.\footnote{From {\it Capitolo contro il portar la toga}
in "Dialogo sopra i due Massimi Sistemi del Mondo Tolemaico e Coperniano" (1632).
}"}
(Galileo Galilei)

\section{Introduction}

In this work, we reconsider the problem of finding simple textures, 
which could possibly be linked to some symmetry principle  
or simply embedded in a grand unification context.
In practice, we have to write two quark mass matrices consistent with the ten experimental observables 
(six mass eigenvalues, three angles and one phase of the CKM mixing matrix) 
in terms of as few parameters as possible,
by establishing simple relations between observables to make the model predictive.

Since the Gatto-Sartori-Tonin (GST) relation $\sin\theta_C \approx \sqrt{m_d/m_s}$ 
was pointed out \cite{GST}, 
a lot of mass matrix models generalized it to three generations (see e.g. \cite{Fritext}--\cite{Branco}).
Yet, the more and more precise experimental data have ruled out or disfavored
many interesting fermion mass textures proposed so far. 
For instance, the pioneering ones proposed by Georgi and Jarlskog (GJ) \cite{GJ} 
do not satisfy the $V_{cb}$ constraint since 16 years \cite{RRR} and,
thanks to the increased sensitivities to $|V_{ub}/V_{cb}|$ and $\sin\beta$,
also the more general class of symmetric textures with zeros in the 11,13 and 31 elements 
\cite{HR} has turn out to be disfavored \cite{FriXi01, RRRV, CRS, Raby}.

\begin{figure}[!t]
\vskip . cm
\centerline{ \psfig{file=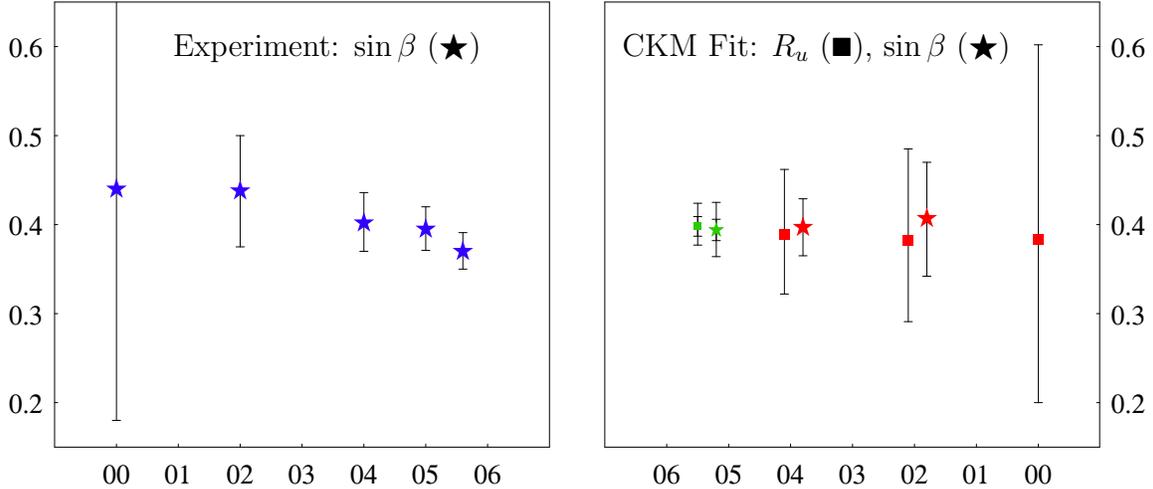,width=1 \textwidth} 
\put(-380, 170){Experiment: $\sin\beta$ ($\bigstar$) } 
\put(-210, 170){CKM Fit: $R_u$ (\footnotesize$\blacksquare$\normalsize), $\sin\beta$ ($\bigstar$)} }
\caption{Left: experimental value of $\sin\beta$ \cite{PDGs}:
average measurements from BaBar and Belle since 2002;
CDF results for year 2000. 
Right: CKM fit value of $R_u$ and $\sin\beta$ according to: 
the Particle Data Group (PDG) \cite{PDGs} at 90\% c.l. from 2000 to 2004; 
CKMfitter Collaboration \cite{ckmf} at 1 and 2$\sigma$ after summer 2005.}
\label{fig1}
\vskip . cm
\end{figure}

Recently, an interesting phenomenological fact has been emerging, 
whose impact on fermion mass matrices deserves proper consideration.
As suggested both from direct measurements \cite{alfadirect} and global CKM fits 
\cite{ckmf}\footnote{As the results of the fits 
carried out by different collaborations \cite{ckmf, fits} agree, 
in this work we adopt for definiteness the fit of ref. \cite{ckmf}.},
the unitarity triangle is rectangle or nearly so: $\alpha\sim 90^\circ$.
Since  $~\sin \alpha = \sin \beta/R_u$,  this can be easily realized for global fits
by comparing the smallest side of the unitarity triangle, $R_u = |\frac{V_{ud} V_{ub}}{V_{cd} V_{cb}}|$,  
with the precisely measured angle in front of it, $\beta$. 
We draw such a comparison in fig. \ref{fig1}, 
also showing how much the robustness of $\alpha \sim 90^\circ$ has increased 
during the last years.

After introducing our notations and conventions in Section 2,
we analyse in Section 3 the simplest possibility that near-neighbor-generation 
mixings are given in terms of the four ratios of quark masses,
while far-neighbor-generation mixings vanish. 
A very important role is played by the non-removable phases between left-chirality families, 
which are a potential source of CP violation (CPV) \cite{HR, FriXi97, Branco, Isamax}. 
To avoid to introduce new parameters, the simplest approach is to assume that 
these phases are either $0,\pm \pi/2$ or $\pi$,  
namely that such sources of CPV are either maximal or switched off.  
It is known that 
the presence of a maximal phase between two quarks or two leptons of left-chirality 
has important consequences:
$\sin \theta_C$ is numerically so close to $\sqrt{m_d/m_s}$ 
that a maximal phase between the lightest generations of quarks 
is needed to preserve the GST relation from corrections due to the up-quark sector 
\cite{FriXi97, FriXi01, RRRV};
a maximal phase between the $\mu$ and $\tau$ leptons permits to end up with a maximal atmospheric angle, 
starting from a $\pi/4$ mixing in the $\nu_\mu-\nu_\tau$ and/or $\mu-\tau$ sectors \cite{Isamax}. 
As we are going to argue, a maximal phase between the heaviest generations of quarks 
or between the $e$ and $\mu$ leptons  
is instead not suitable respectively for $|V_{cb}|$ and the solar angle\footnote{For attempts to obtain
the solar angle by exploiting a maximal CPV phase between the $e$ and $\mu$ leptons 
see ref. \cite{FXlast}.}.

Following the guidelines so obtained,
we are lead to write down a simple set of quark mass matrices, eq. (\ref{2text}),
whose elements are real but for a purely imaginary one and simply depend on quark mass ratios. 
The up-quark mass matrix $\mathbf{m_{u}}$ has a double seesaw-like (or Fritzsch \cite{Fritext}) form;
together with $\mathbf{m_{d}}$, these textures can be viewed as a renewed version of the GJ ones. 
This predictive set belongs to the  category of the so-called "5 zero" textures 
and gives a quite reasonable fit of the quark masses and the CKM mixing matrix, 
with the specific prediction $\alpha \approx \pi/2$.
The weak point is that, in order to fit $\beta$ and $|V_{ub}/V_{cb}|$, 
the up-quark mass comes out too large as compared to the chiral perturbation theory (CPT) result \cite{Leut}.
This is the same problem which has been pointed out, though with different approaches, 
in refs. \cite{FriXi01, RRRV, Raby, Branco}.

We then consider introducing additional small parameters 
that could make the agreement with data more satisfactory
and identify the best choice to cure this problem. 
We show that switching on far-neighbor mixings to fit $\beta$ and $|V_{ub}/V_{cb}|$
would correspondingly deteriorate the fit of $\alpha$, which would now be missed by about $2\sigma$.  
This remedy was adopted in refs. \cite{RRRV, Raby, Branco}
(from their perspective, filling the zero in the 31-element of $\mathbf{m_{d}}$),
when the much poorer sensitivity to $\alpha$ could not reveal a conflict.
Driven by the recent data, we abandon that way and rather adopt a remedy 
which does not alter the desirable prediction $\alpha \approx \pi/2$, 
allowing for vanishing far-neighbor mixings. 
It consists in simply giving up the strict dependence 
on mass ratios involving the up-quark mass
by the introduction of a small parameter in the 11-element of $\mathbf{m_{u}}$. 

Inspired by the analysis in Section 3,
in Section 4 we carry out a more general analysis  
by studying quark textures with 10 parameters and 6 possibly relevant phases.
We show how experimental data single out two sets of quark textures with real or imaginary elements
and which depend only on 5 parameters.
The analysis proves that these two sets of textures, 
among which is the one obtained from the educated guess of Section 3,
are stable solutions in the face of experimental data.
Textures are of course 
defined up to unitary transformations that leave the CKM matrix invariant\footnote{But clearly, 
flavour models do depend on the choice of basis since their symmetries do not commute with basis
transformations.},
however only unitary transformations that do not introduce new small parameters
- namely those that correspond to rotation angles and phases which are $0,\pm \pi/2, \pi$ -
would meet our requirements.
Indeed, it turns out that the down-quark textures of the two sets 
are related by a unitary transformation acting on the right-handed down-quarks
which has only one maximal angle and maximal phases. 

The previous results concern fermion mass textures at the electro-weak scale. 
There are two,
non-independent reasons to transpose our analysis to the grand unification scale: 
1) the extension of the analysis to the lepton sector becomes more predictive in a GUT framework; 
2) $\Lambda \geq  M_{GUT}$ is a natural scale to define broken flavour symmetries 
capable to explain the mass textures. 
Of course the data fitting must now take into account the running of the textures 
down to the electro-weak scale, as done in Section 5.
This evolution modifies the imposed relations among the parameters. 
If only the top coupling is large (small $\tan \beta$), 
the agreement with data is fair, but including the
contributions from the bottom and $\tau$ couplings (large $\tan \beta$)
the agreement is further improved.

In Section 6, we embed the proposed textures in an $SU(5)$ GUT framework \cite{SU5original}. 
The Yukawa couplings in the mass matrices cannot all transform as $SU(5)$ singlets, because 
the invariance of the Yukawa matrices would lead to the prediction $m_e^T=m_d$, 
in contrast with the more realistic GJ relations $m_b=m_\tau$, $3 m_s=m_\mu$, $m_d=3m_e$ \cite{GJ}. 
In general these couplings depend on the v.e.v.'s that break $SU(5)$
in a $\underline{24}$ or $\underline{75}$, and correspondingly transform. 
Since our textures are to a large extent a minimal extension of the GJ ones, 
their well known mechanism can be simply incorporated;
the fit of the quark mass matrices of Section 5 
then leads to lepton masses that agree with experimental values to $O(10\%)$, 
the largest deviation concerning $b-\tau$ unification \cite{btauoriginal}. 
At difference of the quark sector, we do not attempt a better fit 
for lepton masses as it would drastically depend on the details 
of the supersymmetric model \cite{drastic} 
and, possibly, on threshold effects due to heavy particles \cite{corrss}.
Since lepton mixings are less sensitive to these details,
they can be qualitatively analysed by using the lepton textures obtained with the simple GJ prescription:
while the first set displays small mixings for both chiralities, 
the second set has a maximal $\mu-\tau$ mixing for the left-handed charged leptons.


\section{Conventions and Preliminaries}

In the basis of the unknown flavour symmetry the Lagrangian is described by
\bea
{\cal L}~ =& -&~ \bar u_R^{T}~ {\mathbf m_u} ~u_L-~ \bar d_R^{T}~ {\mathbf m_d} ~d_L 
                 - \frac{1}{2}~\nu^T ~{\mathbf m_{\nu}}^{eff}~\nu ~-~ \bar e_R^{T}~ {\mathbf m_e} ~e_L ~ \nn \\
&+&~\frac{g}{\sqrt{2}} ~\bar u_L^T ~\gamma^\lambda ~d_L ~W^+_\lambda  ~
 +~\frac{g}{\sqrt{2}} ~\bar e_L^T ~\gamma^\lambda ~\nu ~W^-_\lambda ~+~ \mathrm{h.c.}~~~~~~\label{lagr}\\ \nn\\
 ~~ {\mathbf m_u} = U_{u_R} \hat m_u U_{u_L}^{~\dagger} & ,& 
 {\mathbf m_d} = U_{d_R} \hat m_d U_{d_L}^{~\dagger}~~~,
 ~~{\mathbf m_{\nu}^{eff}}= U_\nu^* \hat m_\nu U_\nu^\dagger ~~~,
~~ {\mathbf m_e} = U_{e_R} \hat m_e U_{e_L}^{~\dagger} ~~~
\label{lagrU}
\eea
where a hat is placed over a diagonal matrix with real positive eigenvalues 
whose order is established conventionally by requiring $|m^2_2 -m^2_3| \ge m^2_2 - m^2_1 \ge 0$,
and the $U$'s are unitary matrices. 
The CKM and MNS mixing matrices are respectively 
\beq
V_{CKM}=U_{u_L}^{~\dagger} ~{U_{d_L}}~~~~,~~~~U_{MNS}=U_{e_L}^{~\dagger} ~U_\nu~~~.
\eeq

Diagonalization proceeds, at left and right, alternating rotations and 
phase transformations. In particular, we introduce
\bea
R_{23}(\theta_{23})=\footnotesize \left( \matrix{1 & 0& 0 \cr 0 & c_{23} & s_{23} \cr
     0 &  - s_{23} & c_{23}}\right)\normalsize,~
R_{13}(\theta_{13} )=\footnotesize\left( \matrix{ c_{13}& 0& s_{13} \cr 0 & 1 & 0 \cr
      - s_{13} & 0&c_{13}}\right)\normalsize,~
R_{12}(\theta_{12} )=\footnotesize\left( \matrix{c_{12} & s_{12}& 0 \cr 
                              - s_{12}    & c_{12} & 0 \cr
                                   0 & 0  & 1}\right)\normalsize,\nn\label{def}\\ \\
\Phi=\mathrm{diag}( \phi_{12}+\phi_{23} , \phi_{23},1 )
                            ~~~~~~,~~~~~~~~~
\Phi'=\mathrm{diag}(\phi'_1,\phi'_2,\phi'_3)
~~~~~~,~~~~~~~~~~~\nn
\eea
where $s_{ij}=\sin\theta_{ij}$, $c_{ij}=\cos\theta_{ij}$. 
We find particularly convenient to parameterize each $U$ appearing in eq. (\ref{lagrU}), 
as well as $V_{CKM}$ and $U_{MNS}$ themselves, 
in terms of a matrix in the standard CKM parameterization $U^{(s)}$ \cite{PDGs}, 
multiplied at left and right by diagonal matrices of phases of the form of $\Phi$ and $\Phi'$. 
Leaving indices understood, 
\beq
U~=~e^{i \Phi}~U^{(s)}~ e^{i \Phi'}~~~~~~,~~~~~~~~~~~~
 U^{(s)}=R_{23}(\theta_{23}) \Gamma_\delta R_{13}(\theta_{13}) \Gamma_\delta^\dagger R_{12}(\theta_{12})
\label{par}
\eeq 
where $\Gamma_\delta =$diag$(1,1,e^{i \delta})$,
angles belong to $[0,\pi/2]$ and phases to $]-\pi, \pi ]$.

The phases $\Phi'$ of charged fermions can be removed 
by means of unitary transformations acting on right-handed fields. 
Defining $\Phi_q=\Phi_{u_L}- \Phi_{d_L}$, $\Phi_\ell=\Phi_{e_L}- \Phi_\nu$,
one then obtains\footnote{$\Phi'_\nu$ solely contributes to Majorana CPV phases.}
\beq
V_{CKM} ~= ~{U^{(s)}_{u_L}}^\dagger~ e^{-i \Phi_q}~U^{(s)}_{d_L} ~~~~~~,~~ ~~
U_{MNS}~ =~ {U^{(s)}_{e_L}}^\dagger ~e^{-i\Phi_\ell}~U^{(s)}_\nu e^{i \Phi'_\nu} ~~~.
\label{2U}
\eeq
The phases $\phi^{q}_{ij}=\phi^{u_L}_{ij}-\phi^{d_L}_{ij}$ and 
$\phi^{\ell}_{ij}=\phi^{e_L}_{ij}-\phi^{\nu}_{ij}$  ($ij=12,23$) 
can be chosen in $]-\pi,\pi] $ and 
represent the phase difference between the $i-$th and the $i+1-$th generation of left-handed
quarks and leptons respectively, 
before shuffling the flavors to go in the mass basis. 
They cannot be removed and, although not individually measurable\footnote{Of course, 
none of the 10 (12) parameters appearing in the r.h.s. of eqs. (\ref{2U}) is measurable.}, 
are a source for CP violation and play a crucial role in the mixing matrices 
- see e.g. \cite{HR, FriXi97, Branco, Isamax}.
Note that low energy CPV in the CKM (MNS) matrix can be generated in the limit where only 
$\phi_{12}^{q(\ell)}$ and/or $\phi_{23}^{q(\ell)}$ are present, 
as well as in the limit where there are just $\delta^{u_L(e_L)}$ and/or $\delta^{d_L(\nu)}$.

In a grand unified theory like $SU(5)$, one expects $m_e \approx m_d^T$, where
the coefficients may be different for the elements coming from a (fundamental or effective) 
$\overline{\underline{45}}$ Higgs representation. 
From $U_{e_L}\approx U_{d_R}^* $, it follows that $\Phi_{e_L} \approx - \Phi_{d_R}$.
The CPV phases of $\Phi_{d_R}$, irrelevant for the quark sector, are thus important for the lepton sector, 
as they explicitly appear in the expression for $U_{MNS}$. 
In such a framework, it is interesting to investigate possible connections between CPV 
in quarks and leptons, as will be done in Section 6.


\section{Guidelines of the Approach}

The goal of the subsequent approach, based on a couple of assumptions, 
is to establish possible relations among ratios of quark masses, angles and phases of $V_{CKM}$, 
at the low energy scale $m_Z$. 
This discussion turns out to provide useful guidelines to be exploited
in Section 4, where a more general analysis will be provided.
Clearly, some observations are already present in the literature, though obtained with different approaches.
We update and combine them from a point of view that allows  
to cleanly identify the weak points and to introduce our proposed solutions.

Firstly, it is well known that the phenomenological relation $|V_{ub}|=O( |V_{us} V_{cb}|)$ 
supports the possibility that $\theta^{u_L}_{13}$ and $\theta^{d_L}_{13}$ are small enough 
to negligibly contribute to $|V_{ub}|$.
Also the associated phases $\delta^{u_L}$ and $\delta^{d_L}$ become irrelevant, so that 
the fundamental CKM parameters appearing in the r.h.s. eq. (\ref{2U}) are reduced from ten to six: 
four near-neighbor generation mixing angles and the two phases of $\Phi_q$
\footnote{Still, there is redundancy with respect to measurable quantities but, 
at difference of previous analysis \cite{HR, FriXi97, FriXi01},
we prefer to carry out the discussion explicitly in terms of these six parameters 
in order to see separately the role played by $\phi^q_{12}$ and $\phi^q_{23}$ in the CKM matrix.}.
Then, a non-vanishing $|V_{ub}|$ arises in $V_{CKM}$ as the result 
of re-arranging $R^{u_L T}_{12}$ at the right of $R^{u_L T}_{23} R^{d_L}_{23}$.     
Explicitly, from eq. (\ref{2U}) one gets:
\bea
|V_{ub}|~=~ s^{u_L}_{12}~ |s^{d_L}_{23} c^{u_L}_{23} e^{-i\phi_{23}^q}-  s^{u_L}_{23} c^{d_L}_{23}|
           ~=~\frac{s^{u_L}_{12}}{s^{d_L}_{12}}~|V_{td}|~~~~~~,~~~~~~\\
|V_{cb}|~=~\frac{c^{u_L}_{12}}{s^{u_L}_{12}}~|V_{ub}|~~~~~,~~~~~~~
|V_{ts}|~=~\frac{c^{d_L}_{12}}{s^{d_L}_{12}}~|V_{td}|~~~~,~~~~~~~~~~~~~~\label{2ratios}\\\nn\\
|V_{us}|~=~|s^{d_L}_{12} c^{u_L}_{12}e^{-i\phi_{12}^q}-  c^{u_L}_{23} s^{u_L}_{12} c^{d_L}_{23} c^{d_L}_{12}
-s^{u_L}_{23} s^{u_L}_{12} s^{d_L}_{23} c^{d_L}_{12} e^{i\phi_{23}^q}|~~~~~~.
\label{vus} 
\eea
Analogous expressions hold for the corresponding elements of $U_{MNS}$ if one attributes 
the smallness of $|U_{e3}|$ to the individual smallness of $\theta^{e_L}_{13}$ and $\theta^{\nu}_{13}$. 
In such a case, due to $|U_{\mu 3}|\approx 1/\sqrt{2}$, one also has 
$s^{e_L}_{12} \lesssim |U_{e3}|$.

\begin{figure}[!t]
\vskip 0. cm
\centerline{ \psfig{file=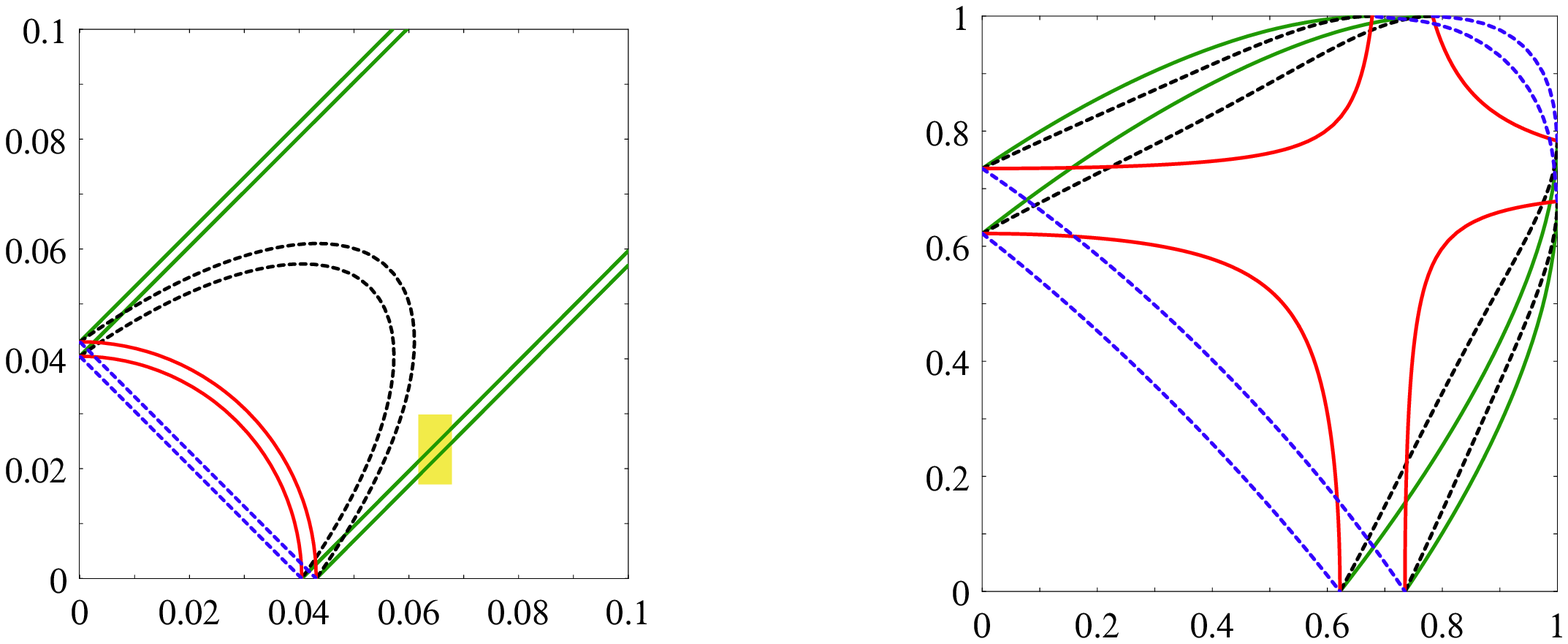,width=1.05 \textwidth} 
\put(-310, -10){\large $s^{u_L}_{23}$} \put(-490, 150){\large $s^{d_L}_{23}$}
\put(-50, -10){\large $s^\nu_{23}$} \put(-230, 150){\large $s^{e_L}_{23}$}
\Green  \put(-386, 140){ $0$} \put(-304, 80){$0$}
\Black  \put(-350, 114){$\pm \frac{\pi}{4}$}
\BrightRed  \put(-388, 72){$\pm \frac{\pi}{2}$}
\Blue  \put(-410, 45){$\pi$}
\Green  \put(-160, 165){$0$} \put(-38, 45){$0$}
\Black  \put(-125, 154){$\pm \frac{\pi}{4}$} \put(-50, 75){$\pm \frac{\pi}{4}$}
\BrightRed  \put(-90, 115){ $\pm \frac{\pi}{2}$}
\Blue  \put(-115, 70){$\pi$}  \put(-30, 170){$\pi$}
\Black}
\caption{Left: stripes consistent with $|V_{cb}|$ at 2$\sigma$ \cite{ckmf} in the plane 
$\{s^{u_L}_{23},s^{d_L}_{23}\}$ for $\phi^q_{23}=0,\pm \pi/4,\pm\pi/2,\pi$.
The shaded rectangular region corresponds to the identifications 
$s^{u_L}_{23} = \sqrt{m_c/m_t}$ and $s^{d_L}_{23}= m_s/m_b$ at $m_Z$,
according to the PDG mass estimates \cite{PDGs}.
Right: stripes consistent with $|U_{\mu3}|$ at 1$\sigma$ \cite{Fogli} in the plane
$\{s^{\nu}_{23},s^{e_L}_{23}\}$ for $\phi^\ell_{23}=0,\pm \pi/4,\pm\pi/2,\pi$.}
\label{fig2}
\vskip 0.2 cm
\vskip .5 cm
\centerline{ \psfig{file=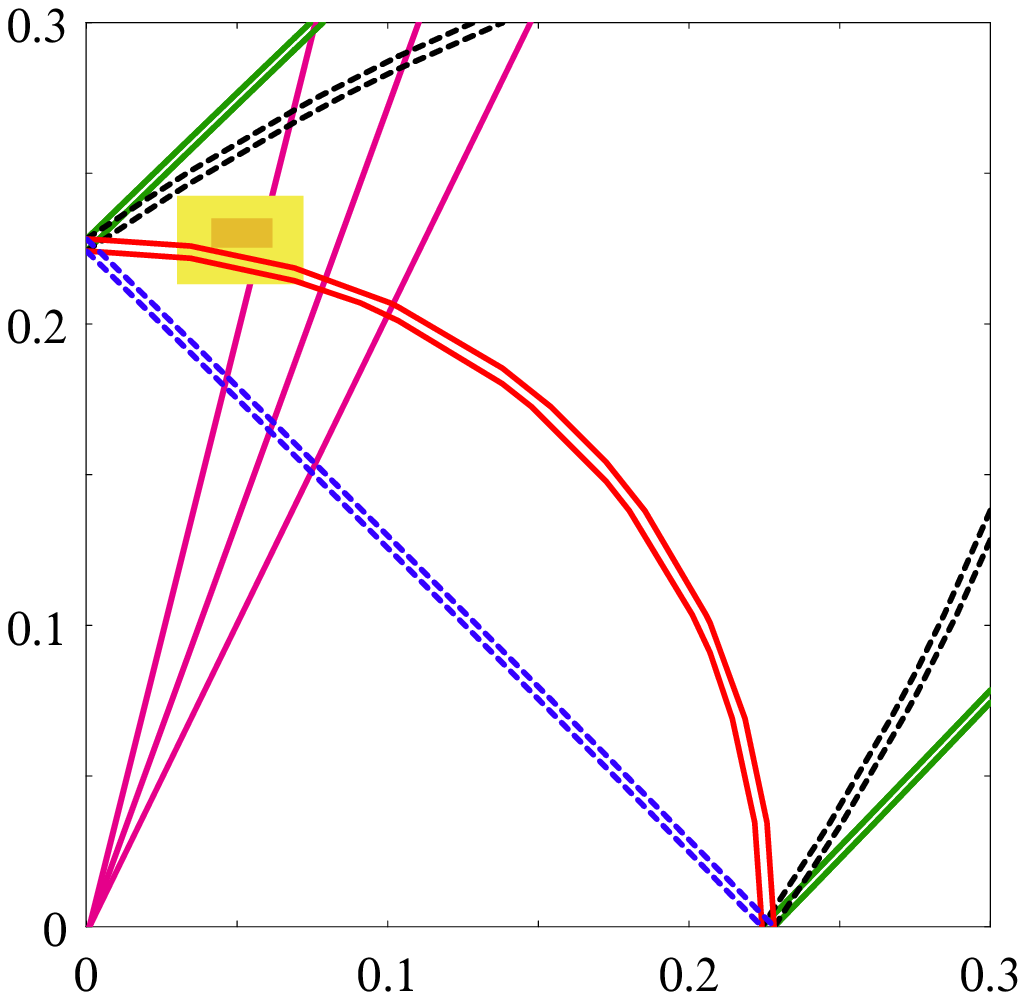,width=.4 \textwidth} 
\put(-50, -10){\large $s^{u_L}_{12}$} \put(-195, 140){\large $s^{d_L}_{12}$}
\Green  \put(-153, 162){ $0$}  \put(-20, 30){ $0$}
\Black  \put(-35, 70){$\pm \frac{\pi}{4}$}
\BrightRed  \put(-105, 125){$\pm \frac{\pi}{2}$}
\Blue  \put(-100, 52){$\pi$}
\SetColor{Red}\BrightRed \SetWidth{1}  
\Line(-166,15)(-125,180) \Line(-166,15)(-105.5,180) \Line(-166,15)(-85,180) 
\Text(-162,58)[c]{$\beta$}\Text(-92,190)[c]{$14^\circ~~ 20^\circ~~26^\circ~~~$}
\CArc(-166,15)(30,75,90)
\Black}
\caption{
Stripes consistent with $|V_{us}|$ at 2$\sigma$ \cite{ckmf} in the plane 
$\{s^{u_L}_{12},s^{d_L}_{12}\}$ for $\phi^q_{12}=0,\pm \pi/4,\pm\pi/2,\pi$.
The larger (yellow) and smaller (orange) shaded rectangular regions correspond 
to the identifications 
$s^{u_L}_{12}= \sqrt{m_u/m_c}$, $s^{d_L}_{12}= \sqrt{m_d/m_s}$ at $m_Z$, 
respectively according to the PDG \cite{PDGs} and to the CPT mass estimates \cite{Leut}. 
For $\phi^q_{12}=+\pi/2$, the straight lines show the value assumed by $\beta$ - see eq. (\ref{eqab}).}
\label{fig3}
\vskip 0. cm
\end{figure}

Secondly, the smallness of the CKM mixing angles suggests that, 
to avoid large cancellations between $U_{u_L}$ and $U_{d_L}$, 
their mixings should be individually small.
In the $c^{u_L}_{12}=c^{e_L}_{12}=1$ approximations, one obtains the two formally similar expressions:
\beq
|V_{cb}| \simeq   |s^{d_L}_{23} c^{u_L}_{23} -  e^{-i\phi_{23}^q}s^{u_L}_{23} c^{d_L}_{23}|~~~~~~~,~~~~~~~
|U_{\mu3}|\simeq |s^\nu_{23} c^{e_L}_{23} -  e^{-i\phi_{23}^\ell} s^{e_L}_{23} c^\nu_{23}|~~.
\label{vcbs}
\eeq
Since the experimental range in the l.h.s. is different, different regions 
in the domains $\{s^{u_L}_{23},s^{d_L}_{23},\phi_{23}^q\}$ and $\{s^\nu_{23},s^{e_L}_{23},\phi_{23}^\ell\}$ 
are allowed, as displayed in fig. \ref{fig2} for 
$\phi_{23}^{q,\ell}=0$, $\pm\pi/4$, $\pm\pi/2$, $\pi$.  
On the quark side, $s^{u_L}_{23}$ and $s^{d_L}_{23}$ have to be small in order to avoid tunings, as just mentioned. 
Notice that the identifications
$s^{u_L}_{23} = \sqrt{m_c/m_t}$ and $s^{d_L}_{23}= m_s/m_b$, 
corresponding to the shaded rectangular region, suggest $\phi^q_{23}\approx 0$. 
On the lepton side, the possibility $\phi^\ell_{23} =\pm \pi/2$ turns out to be very attractive: 
if $\theta^{e_L}_{23}=\pi/4$, 
a maximal atmospheric angle is obtained for whatever value of $\theta^\nu_{23}$,
and viceversa.

Neglecting terms of second order in $s^{u_L}_{23}$ and $s^{d_L}_{23}$, the expressions for $|V_{us}|$ 
and the angles of the unitarity triangle become easy to handle\footnote{We 
do not write the analogous expressions for the lepton sector as the large atmospheric mixing prevent us from
taking small $s^{e_L}_{23}$ and $s^{\nu}_{23}$.}:
\beq
e^{i\beta} ~|V_{us}| \simeq   s^{d_L}_{12} c^{u_L}_{12} -  e^{-i\phi_{12}^q} s^{u_L}_{12}  c^{d_L}_{12}~~~~~~~~,~~~~~~~
\alpha \simeq  \phi_{12}^q  
-\arg(1+ e^{i \phi_{12}^q}\frac{ s^{u_L}_{12} s^{d_L}_{12}}{c^{u_L}_{12} c^{d_L}_{12}})~~~~.
\label{bvusa}
\eeq
It turns out that 
{\it the sole source of CP violation in $V_{CKM}$ is $\phi^q_{12}$},
as a dependence on $\phi^q_{23}$ in eq. (\ref{bvusa}) would by introduced by negligible terms proportional 
to $s^{u_L}_{23} s^{d_L}_{23}$.
The above expression for $|V_{us}|$, formally similar to those in eq. (\ref{vcbs}), 
selects appropriate regions in the domain $\{s^{u_L}_{12},s^{d_L}_{12},\phi_{12}^q\}$, 
as displayed in fig. \ref{fig3} for $\phi_{12}^q =0$, $\pm\pi/4$, $\pm\pi/2$, $\pi$.  
In particular, for $\phi^q_{12}=\pm\pi/2$ the circumference of radius $|V_{us}|$ 
in the plane $\{s^{u_L}_{12},s^{d_L}_{12}\}$ is selected, 
which nicely overlaps the broader 
rectangular region consistent with the identifications 
$s^{u_L}_{12}= \sqrt{m_u/m_c}$, $s^{d_L}_{12}= \sqrt{m_d/m_s}$, as given by the PDG \cite{PDGs}.
The overlap would be poor considering instead the smaller 
rectangular region consistent with CPT analysis \cite{Leut}.

Also the fact that the unitarity triangle is at present compatible with rectangle 
suggests the remarkable value $\phi_{12}^q=+\pi/2$, in agreement with the above discussion on $|V_{us}|$
but solving in addition the sign ambiguity.
Taking $\phi_{12}^q=\pi/2$ and neglecting terms of second order in $s^{u_L}_{12}$ and $s^{d_L}_{12}$, 
one has simply\footnote{Notice that $\pi/2$ is close to the value of $\phi_{12}^q$ that maximizes $\beta$.}
\beq
\alpha \simeq \frac{\pi}{2}  
~~~~~~,~~~~~~~~ 
\tan \beta \simeq \frac{s^{u_L}_{12}}{s^{d_L}_{12}  }  ~~~.
\label{eqab}
\eeq
As shown in fig. \ref{fig3}, the identifications 
$s^{u_L}_{12} = \sqrt{m_u/m_c}$ and $s^{d_L}_{12} = \sqrt{m_d/m_s}$
would point towards relatively small values of $\beta$, 
as compared to the present 1$\sigma$ average of BaBar and Belle: 
$\beta ={21.7^\circ} ^{+1.3^\circ}_{-1.2^\circ}$ \cite{alfadirect}.
Since from eq. (\ref{2ratios}) $s^{u_L}_{12} \simeq |V_{ub}/V_{cb}|$ and $s^{d_L}_{12} \simeq |V_{td}/V_{ts}|$, 
these identifications would also point towards respectively low and large values of 
$|V_{ub}/V_{cb}|$ and $|V_{td}/V_{ts}|$,  
whose $2\sigma$ ranges are $0.087-0.099$ and $0.173-0.224$ according to \cite{ckmf}.

A compromise is reached for 
small $m_d/m_s$ and large $m_u/m_c$:
e.g. $\sqrt{m_d/m_s}=0.22$, $\sqrt{m_u/m_c}=0.08$, in which case $\beta=20^\circ$. 
However, a more accurate analysis reveals a problem: 
since $m_c/m_s \ge 8$ at $m_Z$, one has $m_u/m_d \ge 1$,
in rough contrast with the PDG interval $m_u/m_d=0.5\pm0.2$ \cite{PDGs} 
and, a fortiori, with the CPT prediction, $m_u/m_d=0.553\pm0.043$ \cite{Leut}. 
In summary, a better fit to $\beta$ and $|V_{ub}/V_{cb}|$ would require
a value of $m_u/m_d$ (or equivalently, $m_u/m_c$) which is about three times as much as 
the CPT estimate.

\subsection{Textures for a First Approximation}

Given the delicate problem of the definition of light quark masses, it is nevertheless suggestive
that the experimental data
allow for such a simple interpretation in terms of angles being related to the mentioned ratios of masses
and CPV stemming just from $\phi^q_{12}=\pi/2\simeq\alpha$. 
A pair of textures with all such characteristics is\footnote{As we are going to discuss
in the next section, ${\mathbf m_d}$ could be equivalently replaced by 
\beq
{\mathbf m_d}=\left( \matrix{ 0 & \sqrt{\frac{m_d m_s}{m_b^2}}& 0 \cr
          \frac{1}{\sqrt{2}}\sqrt{\frac{m_d m_s}{m_b^2}}&  i \sqrt{2} \frac{m_s}{m_b} & \frac{i}{\sqrt{2}} \cr
                \frac{i}{\sqrt{2}}\sqrt{\frac{m_d m_s}{m_b^2}}& 0 & \frac{1}{\sqrt{2}}       } \right)
               y_b \frac{v_d}{\sqrt{2}}       ~~~~\nn .      
\label{1text}
\eeq}:
\beq
{\mathbf m_u}=\left( \matrix{ 0 &  \sqrt{\frac{m_u m_c}{m_t^2}}& 0 \cr
          \sqrt{\frac{m_u m_c}{m_t^2}}& 0 & \sqrt{\frac{m_c}{m_t}} \cr
              0 & \sqrt{\frac{m_c}{m_t}} & 1         } \right) y_t \frac{v_u}{\sqrt{2}}~~~,~~~~~~~
{\mathbf m_d}=\left( \matrix{ 0 & \sqrt{\frac{m_d m_s}{m_b^2}}& 0 \cr
           \sqrt{\frac{m_d m_s}{m_b^2}}& i ~\frac{m_s}{m_b} & 0  \cr
             0  &  \frac{m_s}{m_b}& 1         } \right)
               y_b \frac{v_d}{\sqrt{2}} ~~~. 
\label{2text}
\eeq
There are two small parameters in each texture, that induce only near-neighbor mixings 
and determine the mass ratios in the corresponding sector. 
The unique maximal source of CP violation is the purely imaginary 22-element of ${\mathbf m_d}$,  
which leads to 
\beq
\alpha\approx\frac{\pi}{2} ~~~~,~~~~~~\tan\beta\approx \sqrt{\frac{m_u m_s}{m_d m_c}}~~.
\eeq
In summary, 8 observable quantities (1 phase, 3 angles, 4 mass ratios) uncorrelated in the SM 
are not badly reproduced in terms of textures having 4 parameters and real elements, 
but for a purely imaginary one. 
As far as we know, these particular textures have not been emphasized previously.
Notice that they are a renewed version of the GJ ones (where 
${\mathbf m_d}_{32}=0$ and it was premature to address the issue of CPV), with the 
imaginary element being the Yukawa coupling associated to the $SU(5)$-breaking v.e.v. responsible
for the difference in the spectrum of down-quarks and charged leptons (more on this later). 

Since taking ${\mathbf m_d}_{23}=m_s/m_b$ in eq. (\ref{2text}) has negligible impact 
on the previous discussion, 
in practice these textures belong to the category of hierarchical symmetric textures 
with zeros in the 11,13,31 elements, studied in detail in refs. \cite{HR, RRRV, Raby}.
There, it was shown that their general predictions \cite{HR}
$|V_{ub}/V_{cb}|\approx \sqrt{m_u/m_c}$ and $\tan\beta \lesssim \sqrt{m_u m_s/ m_c m_d}$ 
are barely compatible with experiment - see respectively ref. \cite{RRRV} and \cite{Raby}. 
Even though data have changed in the meantime (in particular $\beta$ is now slightly smaller, 
as can be seen in fig. \ref{fig1}) these conclusions remain valid.

It is worth to signal immediately a possible remedy,
which we are going to exploit in the following section.
We already discussed that by taking $s_{12}^{u_L} \approx  \sqrt{3 m_u/m_c}$ 
while keeping the other identifications of angles with mass ratios, 
one properly enhances $\beta$ and $|V_{ub}/V_{cb}|$ without affecting $\alpha \simeq  \pi/2$. 
This factor of 3 could be induced through a slight modification of $\mathbf m_u$, namely 
by filling the zero in the 11-element with a parameter smaller than those already introduced.

\subsection{On the Effect of $\theta_{13}^{u_L,d_L}\neq 0$} \label{nonzero}

It could seem that $s_{13}^{u_L}\neq 0$ and/or $s_{13}^{d_L}\neq 0$ 
help in solving the problem with $\beta$ and $|V_{ub}/V_{cb}|$. 
Instead we show here that the present data on $\alpha$ 
disfavor this possibility. 
For the textures in eq. (\ref{2text}), this modification would be realized 
e.g. by filling the zero in ${\mathbf m_u}_{31}$ and/or ${\mathbf m_d}_{31}$, but we prefer to 
discuss the argument from a more general point of view.   

Assuming that left-chirality quark mixings are small, 
the expression for $|V_{us}|$ in eq. (\ref{bvusa}) is practically unaffected, 
and so is $|V_{cb}|$ in eq. (\ref{vcbs}) if $s^{d_L}_{13},s^{u_L}_{13} \ll |V_{cb}|/s^{u_L}_{12}$. 
On the contrary, defining 
\beq
e^{i\phi_{cb}}|V_{cb}| \simeq -s^{d_L}_{23} c^{u_L}_{23}  e^{-i\phi_{23}^q}+s^{u_L}_{23} c^{d_L}_{23}~~~,~~~~ 
e^{i\phi_{us}}|V_{us}| \simeq s^{d_L}_{12} c^{u_L}_{12} - e^{-i\phi_{12}^q} s^{u_L}_{12} c^{d_L}_{12}~,
\eeq
and neglecting for simplicity the terms of second order in $s^{u_L}_{12}$ and $s^{d_L}_{12}$:
\bea
e^{-i \xi_b }\left|\frac{V_{ub}}{V_{cb}}\right| \simeq~ \frac{s^{u_L}_{12}}{c^{u_L}_{12}} 
                    + \frac{s^{d_L}_{13}}{|V_{cb}|} e^{i(\delta^{d_L}+\phi_{12}^q+\phi_{23}^q+\phi_{cb})} 
                    - \frac{s^{u_L}_{13}}{|V_{cb}|} e^{i(\delta^{u_L}+\phi_{cb})}~~~,  \nn\\
e^{-i \xi_t}\left|\frac{V_{td}}{V_{ts}}\right| \simeq~  \frac{s^{d_L}_{12}}{c^{d_L}_{12}}  
                    + \frac{s^{d_L}_{13}}{|V_{cb}|} e^{i(\delta^{d_L}+\phi_{23}^q+\phi_{cb})}  
                    - \frac{s^{u_L}_{13}}{|V_{cb}|} e^{i(\delta^{u_L}-\phi_{12}^q+\phi_{cb})}~~~,  
                    \label{gencorr}\\\nn\\
\alpha  \simeq  \phi^q_{12} - \xi_t +\xi_b ~~~~~,~~~~~~\beta   \simeq  \phi_{us} +\xi_t~~~~~.~~~~~~~~~~~~~~ \nn 
\eea
For $s^{d_L(u_L)}_{13}$ to be effective in correcting $|V_{ub}/V_{cb}|$, one needs
$s^{d_L(u_L)}_{13} \approx  s^{u_L}_{12} |V_{cb}|$, in agreement with the hypothesis that $|V_{cb}|$
is slightly affected.

We now take $\phi_{12}^q=\pi/2$, $\phi_{23}^q=0$ (hence $\phi_{cb}=0$)
and consider the effect of $s^{d_L(u_L)}_{13}$ 
only.
As one can see from eqs. (\ref{gencorr}), $|V_{ub}/V_{cb}|$ is maximized
for $\delta^{d_L}=-\pi/2$ ($\delta^{u_L}=\pi$) while, at the same time, 
$|V_{td}/V_{ts}|$ is negligibly affected, $\xi_b$ vanishes, 
$\tan \xi_t= \frac{c^{d_L}_{12}}{ s^{d_L}_{12}|V_{cb}|}s^{d_L(u_L)}_{13} ~$ and
\beq
\alpha  \simeq  \frac{\pi}{2} - \xi_t  ~~~~~,~~~~~~~~~
\beta   \simeq  \arctan (\frac{s^{u_L}_{12}}{s^{d_L}_{12}})   +\xi_t  ~~.
\eeq
With respect to their values in eq. (\ref{eqab}), 
$\beta$ and $\alpha$ are respectively enhanced and lowered by $\xi_t$. 
Fig. \ref{fig3} shows that, in order to fit $\beta$ and $|V_{ub}/V_{cb}|$ 
while keeping the identifications $s^{u_L}_{12}=\sqrt{m_u/m_c}$ and $s^{d_L}_{12}=\sqrt{m_d/m_s}$, 
one needs $\xi_t\sim8^\circ$. However, {\it all this happens 
at the price of turning $\alpha$ down to $82^\circ$, close to its 2$\sigma$ lower range}.

This remedy was implemented for hierarchical symmetric textures with zeros in the 11,13,31 elements 
in order to properly enhance their predictions for $\beta$ and $|V_{ub}/V_{cb}|$, 
in particular by filling ${\mathbf m_d}_{31}$ \cite{RRRV, Raby}. 
Also for hermitian mass matrices with vanishing 11-elements,
the importance of the so-called "non-factorizable" phase $\delta^{d_L}$  
to possibly enhance $\beta$ was pointed out \cite{Branco}.
However, the correlation between $\alpha$ and $\beta$ did not emerged
as the constraint on $\alpha$ were much weaker at that time. 
The present inadequacy of this remedy\footnote{Actually, for some values of $s_{13}^{d_L}$ one can find
corresponding values of $\delta^{d_L}$ that give a good fit to both $\alpha$ and $\beta$.
Such values of $\delta^{d_L}$ being not close to zero or $\pi/2$, this amounts to introducing
one more parameter in the texture. We thank A. Romanino for calling our attention to this possibility.} 
prompted us in re-addressing 
the issue of finding simple textures that correctly reproduce the data.

As already mentioned, for the purpose of enhancing $\beta$ and $|V_{ub}/V_{cb}|$
a better remedy that we are going to adopt in the next section 
consists in filling the zero of ${\mathbf m_u}_{11}$ 
to obtain $s_{12}^{u_L} \approx  \sqrt{3 m_u/m_c}$.  
In top of this, one could successfully exploit a non vanishing $s^{d_L(u_L)}_{13}$ 
but with $\delta^{d_L}=\pi$ ($\delta^{u_L}=\pi/2$). 
Indeed, in this case $\xi_t=0$, 
$\tan \xi_b=\frac{c^{u_L}_{12}}{s^{u_L}_{12}|V_{cb}|}s^{d_L(u_L)}_{13} ~$ and
\beq
\alpha  \simeq  \frac{\pi}{2} + \xi_b  ~~~~~,~~~~~~~~~
\beta   \simeq  \arctan (\frac{s^{u_L}_{12}}{s^{d_L}_{12}})~~~~.
\eeq 
Having $\xi_b\sim 8^\circ$ makes $\alpha$ reach its present central value, about $98^\circ$, 
without affecting $\beta$.
Depending on the closeness of $\alpha$ to $\pi/2$,
future experiments will clarify whether such further correction is really needed.
Notice also that, in this case, $|V_{td}/V_{ts}|$ gets lowered  
while $|V_{ub}/V_{cb}|$ is negligibly affected.


\section{The Textures for $\mathbf m_u$ and $\mathbf m_d$}

We now proceed to determine textures defined at $m_Z$ in good agreement with experiment
and fulfilling the conditions identified in the previous section.
The Wolfenstein parameterization of the CKM matrix \cite{Wolf} in terms of powers of 
$\lambda = \sin\theta_C$ 
suggests to do the same for the mass matrices.
We assume that the textures $\mathbf T_u=\frac{\sqrt{2} \mathbf m_u}{y_t v_u}$ 
and $\mathbf T_d=\frac{\sqrt{2}\mathbf m_d}{y_b v_d}$ are 
\beq
\mathbf T_u=\left( \matrix{  e^{i \phi_c} c ~\lambda^8 & a \lambda^6 & 0 \cr
              a\lambda^6 & 0 & b \lambda^2 \cr
              0 & b \lambda^2 & 1         } \right) 
~~~~,~~~~~
\mathbf T_d=\left( \matrix{ < O(\lambda^5) & g~ \lambda^3 & e^{i \phi_n} n~ \lambda^3 \cr
              g' ~\lambda^3 & e^{i \phi_j} j ~\lambda^2 & e^{i \phi_t} t  \cr
              e^{i \phi_m} m ~\lambda^3 & e^{i \phi_k} k~  \lambda^2 & 1         } \right)
              \frac{1}{\sqrt{1+t^2}}~~,
\label{compltext}
\eeq
where $\lambda=0.23$,
$a,b,c,g,g',j,t,m,n, k$ are $O(1)$ real positive numbers, 
some non-physical phases have already been removed
and the remaining ones are chosen for definiteness in $]-\pi,\pi]$.

For up quarks, introducing $r_u = |1+ e^{i\phi_c}\frac{c b^2}{a^2}|$, 
one has at leading order 
\beq
\sqrt{\frac{m_u}{m_c}} \simeq \sqrt{r_u} \frac{a}{b^2}\lambda^2  ~~~~~, ~~~~~~~~~~
\sqrt{\frac{m_c}{m_t}}\simeq b \lambda^2  ~~~~~,
\eeq
\bea
s^{u_L,u_R}_{23} \simeq \sqrt{\frac{m_c}{m_t}}~~~~,~~~~~s^{u_L,u_R}_{13} \simeq O(\lambda^8)~~~~,~~~~
s^{u_L,u_R}_{12} \simeq \sqrt{\frac{1}{r_u}\frac{m_u}{m_c}} ~~~,~~~~\\ \nn \\
\phi_{12}^{u_L,u_R}=\pi~~~~~~~, ~~~~\phi_{23}^{u_L,u_R}=0~~~~.~~~~~~~~~~~~~~~~~~~
\eea

For down quarks, introducing $r_d =  \frac{|g'-e^{i(\phi_m+\phi_t)} m t|}{g \sqrt{1+t^2}}$, 
one has at leading order
\beq
e^{-i\phi^{d_L}_{12}}\sqrt{\frac{m_d}{m_s+m_d}} \simeq
\frac{1}{\sqrt{r_d}}\frac{ g'-e^{i(\phi_m+\phi_t)}m t}{e^{i \phi_j} j-e^{i(\phi_t +\phi_k)}t k}\lambda 
~~~,~~~~~~
e^{-i\phi_{12}^{d_R}}\frac{m_s}{m_b} \simeq
\frac{e^{i \phi_j} j - e^{i (\phi_t +\phi_k)}t k }{1+t^2} \lambda^2  ~~~,
\label{ratios}
\eeq
\bea
s^{d_L}_{23} e^{-i \phi^{d_L}_{23}}&\simeq& \frac{e^{i (\phi_j-\phi_t)} t j + e^{i \phi_k} k }{1+t^2} \lambda^2 ~~~~~~~,~~~
s^{d_R}_{23} e^{-i \phi^{d_R}_{23}}\simeq e^{-i \phi_t} \frac{t}{\sqrt{1+t^2}} ~~~~~,\label{sd23}\\
s^{d_L}_{13} e^{i \delta^{d_L}_{13}}&\simeq& e^{-i \phi_t} \frac{ t g'+e^{i(\phi_m+\phi_t)}m }{1+t^2} \lambda^3~~~~~,~~~
s^{d_R}_{13} e^{i \delta^{d_R}_{13}}\simeq e^{-i \phi_n} \frac{n}{\sqrt{1+t^2}} \lambda^3~~~,\label{s13}\\
s^{d_L}_{12} &\simeq&  \sqrt{r_d \frac{m_d}{m_s+m_d}} ~~~~~~~~~~,~~~~~~~~~~~~~~~~~
s^{d_R}_{12} \simeq \sqrt{\frac{1}{r_d} \frac{m_d}{m_s+m_d}} ~.
\eea

In particular, as for the phases one ends up with
\beq
\phi_{23}^q = -\phi^{d_L}_{23}~~~~,~~~~~
\phi_{12}^q= \pi - \phi^{d_L}_{12}~~~~,~~~~\delta^{d_L}=\delta^{d_L}_{13}-\phi_{23}^q-\phi_{12}^q+\pi~~~.
\eeq

\begin{figure}[!t]
\vskip . cm
\centerline{ \psfig{file=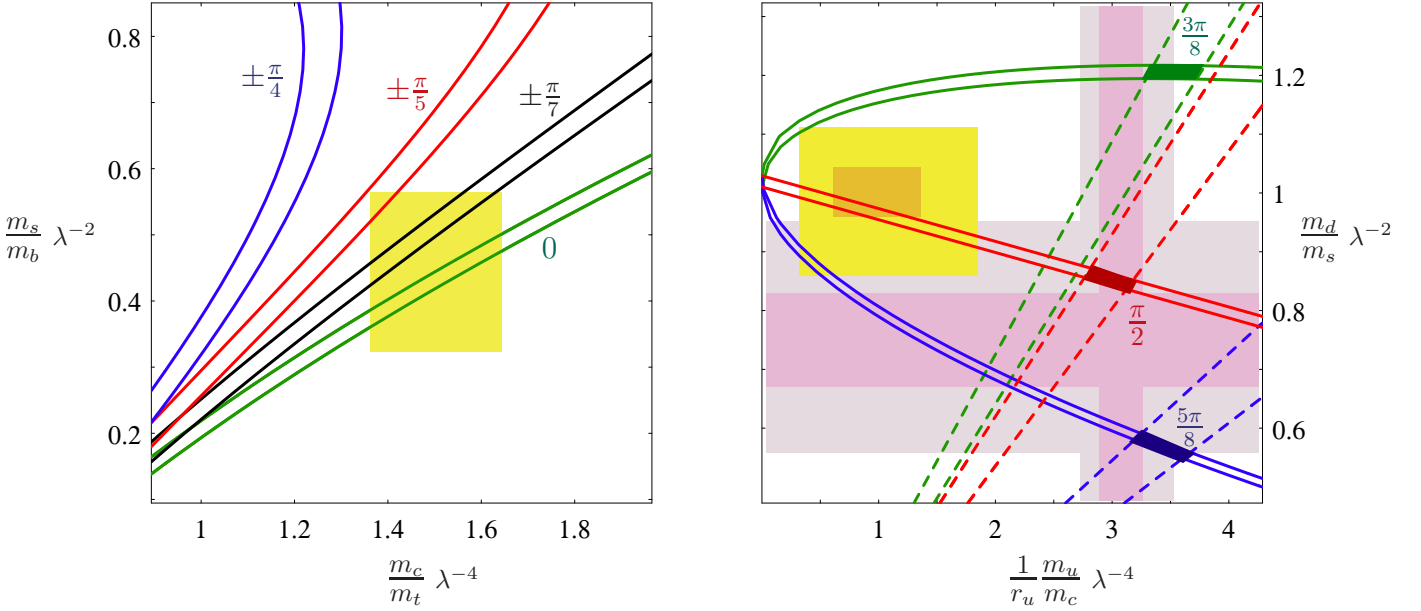,width=1.05 \textwidth} 
\put(-360, -10){\large $\frac{m_c}{m_t}$\footnotesize$~\lambda^{-4}$} 
\put(-505, 120){\large $\frac{m_s}{m_b}$\footnotesize$~\lambda^{-2}$}
\put(-125, -10){\large $\frac{1}{r_u} \frac{m_u}{m_c}$\footnotesize$~\lambda^{-4}$}
\put(-15, 120){\large $\frac{m_d}{m_s}$\footnotesize$~\lambda^{-2}$}
\Green  \put(-305, 115){ $0$} 
\Black  \put(-310, 173){$\pm \frac{\pi}{7}$}
\BrightRed  \put(-360, 175){$\pm \frac{\pi}{5}$}
\Blue  \put(-415, 180){$\pm \frac{\pi}{4}$}
\Green \put(-60, 196){$\frac{3\pi}{8}$} 
\BrightRed  \put(-80, 87){\large $\frac{\pi}{2}$}
\Blue  \put(-62, 48){$\frac{5\pi}{8}$}
\Black}
\caption{
Left: stripes consistent with $|V_{cb}|$ at 1$\sigma$ \cite{ckmf} 
for $\phi_{23}^q =  0,\pm \pi/7,\pm \pi/5,\pm  \pi/4$. 
The shaded rectangle is the range for the mass ratios according to PDG \cite{PDGs}.
Right: stripes consistent with $|V_{us}|$ (full) and $\beta$ (dashed) at 1$\sigma$
for $\phi_{12}^q =  5\pi/8$, $\pi/2$, $3\pi/8$,
with intersections emphasized.
The shaded vertical and horizontal bands signal the values of
$|V_{ub}/V_{cb}|$ and $|V_{td}/V_{ts}|$ at 1,2$\sigma$. 
The larger and smaller shaded rectangles display the range 
of the mass ratios ($r_u=1$) according to PDG \cite{PDGs} and CPT estimates \cite{Leut} respectively.}
\label{fig4}
\vskip 1. cm
\end{figure}

Because the mixing angles are often better known than the masses,
we now express the CKM elements as a function of the quark mass ratios. 
We already know from the previous discussion that, having $s^{u_L}_{23}\simeq\sqrt{m_c/m_t}$, 
it is desirable to also have $s^{d_L}_{23}\simeq m_s/m_b$. 
From eq. (\ref{sd23}), one realizes that this can be achieved naturally in three cases,
which correspond to put a zero respectively in the 23,32,22 element of $\mathbf T_{d}$:
1) $t=0$, $k=j~;$   
2) $t=1$, $k=0~;$  
3) $t=1$, $j=0~.$  
Since $\lambda^3\sim s^{d_L}_{13}  \ll |V_{cb}|/s^{u_L}_{12}\sim 1$, in all these cases 

\beq
|V_{cb}| \simeq \left| \sqrt{\frac{m_c}{m_t}}-  e^{-i \phi_{23}^q}\frac{m_s}{m_b}  \right| 
\eeq

\noindent and, as can be seen from fig. \ref{fig4}, small values of $\phi_{23}^q$ are favored.
According to the various cases, the condition $\phi_{23}^q=0$ reads:
1) $\phi_k=0$ ;~ 
2) $\phi_j=\phi_t$ ;~  
3) $\phi_k=0$~~.\vskip .2cm

Having $s^{u_L}_{12}=O(\sqrt{m_u/m_c})$, 
we also know that it would be convenient to enforce $s^{d_L}_{12}=\sqrt{m_d/m_s}$, 
namely $r_d=1$, which is realized in the various cases for:
1) $~g'=g~;$ 
2)\&3) $~g \sqrt{2}=g_{||}$ , where $e^{i \gamma_{||}} g_{||} = g'-e^{i (\phi_m+\phi_t)} m~.$
In all these cases then 

\beq
|V_{us}|\simeq \sqrt{\frac{m_s}{m_d+m_s}}~ 
\left| \sqrt{\frac{m_d}{m_s}} - e^{-i \phi_{12}^q}\sqrt{\frac{1}{r_u}\frac{m_u}{m_c}} \right| 
\eeq

\noindent and, as can be seen from fig. \ref{fig4}, the experimental value of $|V_{us}|$ favors 
$\phi_{12}^q$ close to $\pm \frac{\pi}{2}$. For the three cases, this would imply:
1) $\phi_j \approx \pm \frac{\pi}{2}$ ;~
2) $\phi_j\approx \pm \frac{\pi}{2}+\gamma_{||}$ ;~  
3) $\phi_k \approx \mp \frac{\pi}{2}+\gamma_{||}$~~.\vskip .2cm

Using these results, we now turn to consider the effect of $s^{d_L}_{13}$. 
By adapting eq. (\ref{s13}) one has:
1) $s^{d_L}_{13}\simeq\frac{m}{g} ~\frac{m_s}{m_b}  \sqrt{\frac{m_d}{m_s}}$ ;~ 
2)\&3) $s^{d_L}_{13}\simeq \frac{ g_{\perp}}{g_{||}} ~\frac{m_s}{m_b}  \sqrt{\frac{m_d}{m_s}}$ ,
where $g_{\perp} = |g'+e^{i(\phi_m+\phi_t)} m|$~.\\ 
In particular, it turns out that the requirement $~s^{d_L}_{13}=0~$ is fulfilled when:
1) $m=0~;$  
2)\&3) $g_\perp=0$, 
i.e. $\phi_t+\phi_m=\pi$ and $g'=m$ , so that also $\gamma_{||}=0$ .   
Then, following the discussion in the previous section, one finds the simple relations  
\bea
\left| \frac{V_{ub}}{V_{cb}} \right|^2 \simeq \frac{1}{r_u} \frac{m_u}{m_c}~~~~~,~~~~~~~~~~~~~~
\left| \frac{V_{td}}{V_{ts}} \right|^2 \simeq \frac{m_d}{m_s}~~~~,~~~~~~~~\nn \\  \\
\alpha  \simeq  \phi^q_{12}~~~~~,~~~ ~~~~
\beta   \simeq  \arg( \sqrt{\frac{m_d}{m_s}}-e^{-i\phi^q_{12}} \sqrt{\frac{1}{r_u}\frac{m_u}{m_c}} )~~.\nn
\eea

\noindent As can be checked from fig. \ref{fig4}, to satisfy all these constraints
it is better to have $\phi^q_{12}=+\frac{\pi}{2}$. The CPT estimate
then suggests $r_u \approx 1/3$.
This is realized for instance with $\phi_c =\pi$ and
$c b^2/a^2 \approx 2/3$ or $4/3$.
Notice that, for the various cases, the condition $\phi^q_{12}=+\frac{\pi}{2}$ reads: 
1)\&2) $\phi_j=\frac{\pi}{2}$ ;~ 
3) $\phi_k =- \frac{\pi}{2}$, in conflict with the former condition $\phi_{23}^q=\phi_k=0$~. 
In the following we then abandon case 3).
 
Summarizing the results, 
the data suggest the simple structures\footnote{
We choose $n=0$ although the data only give a limit $n \le O(\lambda^3)$.} 
\bea
\mathbf T_u=\left( \matrix{  -c \lambda^8 & a \lambda^6 & 0 \cr
              a\lambda^6 & 0 & b \lambda^2 \cr
              0 & b \lambda^2 & 1         } \right)~~~~~,~~~~~~
              c\approx\frac{2a^2}{3b^2}~ \mathrm{or} ~\frac{4a^2}{3b^2}~~,~~~~~~~~~\nn\\ \label{ourtext} \\
\mathbf T_d^{1)}=\left( \matrix{ 0 & g \lambda^3 & 0 \cr
             g \lambda^3 & i ~j \lambda^2 & 0  \cr
             0  & j \lambda^2 & 1         } \right)  ~~~~,~~ ~~~    
\mathbf T_d^{2)}=\left( \matrix{ 0 & \hat g \lambda^3 & 0 \cr
              \frac{\hat g}{\sqrt{2}} \lambda^3 & i ~\hat j \lambda^2 & i \cr
                i~\frac{\hat g}{\sqrt{2}} \lambda^3 & 0 & 1         } \right) \frac{1}{\sqrt{2}} \nn
\eea
where $\hat g= \sqrt{2}g$, $\hat j= 2j$ and $r_u \approx 1/3$. 
Notice that $\mathbf T_u$ has two parameters and a suitable zero filling in the 11 element.
In both cases $\mathbf T_d$ has two parameters, its elements are either real or purely imaginary. 
These sets of textures are thus predictive\footnote{Strictly speaking there are two prediction,
since eight observables (4 mass ratios and the 3+1 parameters of the CKM)
are reproduced starting from five real numbers and one maximal phase.} 
and induce in particular
\beq
\alpha\simeq \frac{\pi}{2}~~~~, ~~~~~~~
\tan \beta   \simeq  \sqrt{ \frac{1}{r_u}\frac{m_u}{m_c}\frac{m_s}{m_d}}~~.
\label{ouraeb}
\eeq 

It is immediate to realize that $\mathbf T_d^{1)}$ and $\mathbf T_d^{2)}$ 
differ by a unitary matrix containing $R_{23}(\theta_{23}=\pi/4)$: 
\beq
\mathbf T_d^{2)}=\left( \matrix{ 1 & 0 & 0 \cr
                                0 & \frac{1}{\sqrt{2}} & \frac{i}{\sqrt{2}}  \cr
                                0 & \frac{i}{\sqrt{2}}  & \frac{1}{\sqrt{2}}   } \right) \mathbf T_d^{1)}~~.
\label{relaz}
\eeq   

\noindent This is reflected in the magnitude of the angle $\theta^{d_R}_{23}$, which is
respectively zero and maximal,
while the remaining non-vanishing angles are directly related to ratios of eigenvalues
\bea
1)~~ s^{d_R}_{23}=0~~~~~~~~~~~~~2)~ s^{d_R}_{23}=\frac{1}{\sqrt{2}}~~~~~~,~~~~~~~~~~~~~~~~~~~~\\
s^{u_L,u_R}_{23} \simeq \sqrt{\frac{m_c}{m_t}}~~,~~~~~~
s^{d_L}_{23}\simeq\frac{m_s}{m_b}~~,~~~~~~
s^{u_L,u_R}_{12} \simeq \sqrt{\frac{1}{r_u}\frac{m_u}{m_c}}~~,~~~~~~
s^{d_L,d_R}_{12}\simeq\sqrt{\frac{m_d}{m_s}}~~.
\eea

\begin{figure}[!t]
\vskip 2. cm
\centerline{ \psfig{file=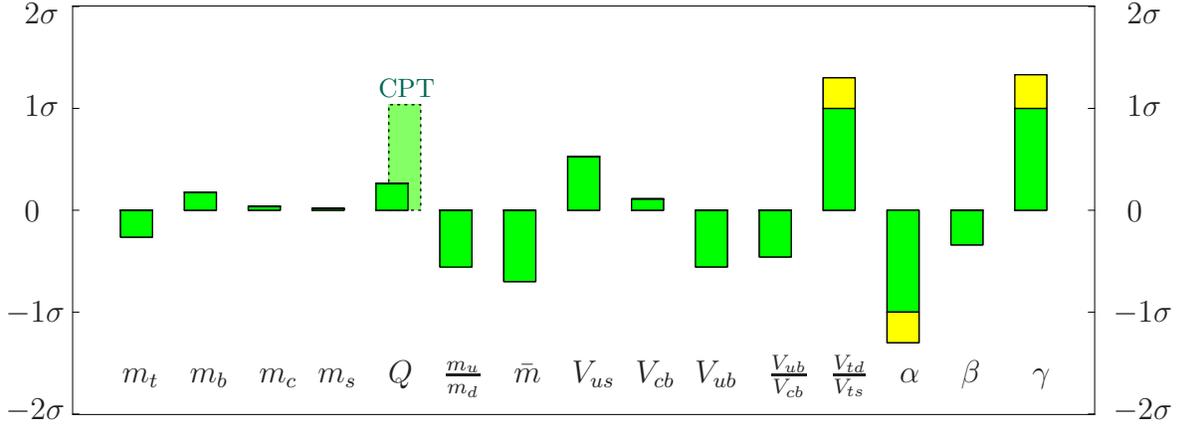,width=.85 \textwidth} 
\Green \put(-275, 120){\footnotesize CPT} \Black
\put(-372, 12){$m_t~~~m_b~~~m_c~~m_s~~~Q~~~\frac{m_u}{m_d}~~~\bar m~~~V_{us}~~V_{cb}~~V_{ub}
~~~\frac{V_{ub}}{V_{cb}}~~\frac{V_{td}}{V_{ts}}~~~\alpha~~~~\beta~~~~~\gamma$}
\put(-416, -2){$-2\sigma$}\put(3, -2){$-2\sigma$}
\put(-416, 35){$-1\sigma$}\put(3, 35){$-1\sigma$}
\put(-409, 73){$0$}\put(8, 73){$0$}
\put(-409, 113){$1\sigma$}\put(8, 113){$1\sigma$}
\put(-409, 148){$2\sigma$}\put(8, 148){$2\sigma$}}
\caption{Predictions of textures in eq. (\ref{ourtext}) with $y_t=0.99$, $y_b=0.17$ ($\tan \beta=10$)
and the coefficients of eq. (\ref{fit1}).
For quark masses we use the PDG intervals \cite{PDGs} while for CKM we adopt the fit of \cite{ckmf} 
- see appendix for details. Explicitly, we obtain $Q=23.5$, 
consistent within 1$\sigma$ with the result of CPT \cite{Leut} (dashed bar).   }
\label{figfit1}
\vskip 1.5 cm
\end{figure}

From fig. \ref{fig4}, it is straightforward to fit the well known CKM angles and phase to 
predict the less known mass ratios. Although the up, down and strange quark
masses are not known with great accuracy, there is one relation 
among the three masses \cite{ilQ} which is accurately known, 
$1 = \frac{1}{Q^2} \frac{m_s^2}{m_d^2} + \frac{m_u^2}{m_d^2}$.
Calculations based on CPT \cite{Leut} find for the ellipse parameter $Q = 22.7 \pm 0.8$.
A good fit of the textures in (\ref{ourtext}) is achieved with
\bea
\frac{m_s}{m_b} \lambda^{-2} \simeq j = 0.419~~~~,~~~~
\sqrt{\frac{m_d}{m_s}}\frac{m_s}{m_b}  \lambda^{-3} \simeq g =0.41~~~,~~~~~~~~~\nn \\ \label{fit1}\\
\sqrt{\frac{m_c}{m_t}} \lambda^{-2} \simeq  b=1.23~~~,~~~~
\sqrt{\frac{1}{r_u}\frac{m_u}{m_c}}\frac{m_c}{m_t}  \lambda^{-6} \simeq a = 2.58~~~,~~~~c =3.48~. \nn
\eea
To have for instance $\tan \beta=10$ we also choose $y_t=0.99$, $y_b=0.17$.
Fig. \ref{figfit1} allows for a quick check of the goodness of the 
fit (central values and $\sigma$'s are collected in the appendix): 
the listed observables are predicted to depart less than $1\sigma$ 
from their central value, but for $|V_{td}/V_{ts}|$ and $\alpha$ 
- hence also $\gamma$ - which are a little bit beyond. 
In particular, according to both direct measurement \cite{alfadirect} and recent CKM fits \cite{ckmf, fits},
the central value of $\alpha$ is close to $98^\circ$. Future sensitive
measurements of this angle will clarify whether the unitarity triangle
is actually rectangle, thus supporting the textures in eq. (\ref{ourtext})
and their simple way of introducing CP violation. 
 
\begin{figure}[!h]
\vskip 1. cm
\centerline{ \psfig{file=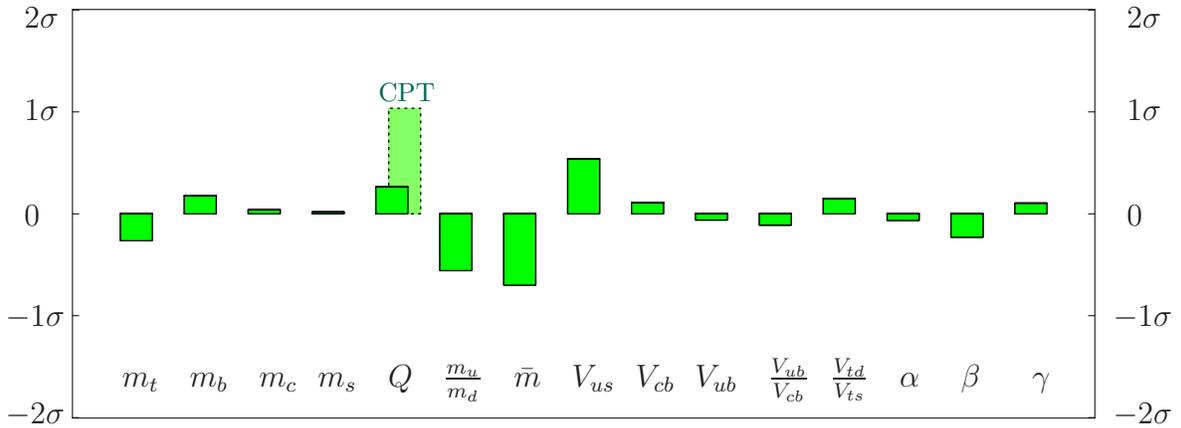,width=.85 \textwidth} 
\Green \put(-275, 120){\footnotesize CPT} \Black
\put(-372, 12){$m_t~~~m_b~~~m_c~~m_s~~~Q~~~\frac{m_u}{m_d}~~~\bar m~~~V_{us}~~V_{cb}~~V_{ub}
~~~\frac{V_{ub}}{V_{cb}}~~\frac{V_{td}}{V_{ts}}~~~\alpha~~~~\beta~~~~~\gamma$}
\put(-416, -2){$-2\sigma$}\put(3, -2){$-2\sigma$}
\put(-416, 35){$-1\sigma$}\put(3, 35){$-1\sigma$}
\put(-409, 73){$0$}\put(8, 73){$0$}
\put(-409, 113){$1\sigma$}\put(8, 113){$1\sigma$}
\put(-409, 148){$2\sigma$}\put(8, 148){$2\sigma$}}
\caption{Same as fig. \ref{figfit1}, but with $e^{i \phi_m} m \lambda^3= i~ 0.05~\lambda^3$ 
as the 31-element of $\mathbf T_{d}^{1)}$.}
\label{figfit2}
\vskip 1.2 cm
\end{figure}

On the contrary, if in future $\alpha$ will turn out to be incompatible with rectangle,
the textures in eq. (\ref{ourtext}) require a proper modification, 
which we now turn to implement through $\mathbf T_d^{1)}$ - but it is straightforward 
to do the same for $\mathbf T_d^{2)}$, given eq. (\ref{relaz}).   
The discussion in Section \ref{nonzero} 
suggests to exploit the correction induced by $s^{d_L}_{13}\neq0$ with $\delta^{d_L}=\pi$
in order to enhance $\alpha$ without altering $\beta$.  
For $\mathbf T_d^{1)}$, these two conditions imply $m \neq 0$ and $\phi_m= \pi/2$.
In this way $\xi_t=0$, $\beta$ is practically unaffected with respect to eq. (\ref{ouraeb}) while
\bea
\alpha  \simeq  \frac{\pi}{2}  + \xi_b ~~~,~~~~~~~~~~~
\tan \xi_b= \frac{m}{g} \frac{m_s/m_b}{\tan\beta|V_{cb}|} \sim \frac{m}{g}~~~,~~~~~~~~~ \nn\\  \\
 \left|\frac{V_{td}}{V_{ts}}\right| \simeq  \sqrt{\frac{m_d}{m_s}} ~(1 - \tan \xi_b \tan \beta ) ~~~~,~~~~~~~~~~
\left| \frac{V_{ub}}{V_{cb}}\right| \simeq \sqrt{\frac{1}{r_u}\frac{m_u}{m_c}} ~| 1-i \tan \xi_b | ~~.~  \nn
\eea

\noindent One needs $\xi_b\sim 9^\circ$, namely $m/g \sim 0.12$ to reach the central value of $\alpha$,
in which case the negative correction to $|V_{td}/V_{ts}|$, $\tan \xi_b \tan \beta \sim 7\%$, 
brings it close to the central value too.
On the contrary, the correction to $| V_{ub}/V_{cb}|$ , $\tan^2\xi_b \sim 2\%$, 
is positive and mild.

For a direct comparison with the fit of fig. \ref{figfit1}, 
we keep the same values as in eq. (\ref{fit1}) with, in addition, 
$~e^{i \phi_m} m \lambda^3= i~ 0.05~\lambda^3$ as the 31-element of $\mathbf T_{d}^{1)}$.
As shown in fig. \ref{figfit2}, the fit to $\alpha$, $\gamma$ and $|V_{td}/V_{ts}|$ 
is further improved, while the other observables do not change significantly.
Notice that, due to the smallness of $m$, the power of $\lambda$ associated to the element $\mathbf T_{d31}^{1)}$
is actually 5, rather then 3. In a flavour theory,
such a small correction to the leading structure is likely to follow as a sub-leading effect.

\vskip .9 cm


\section{Textures at $M_{GUT}$}

We now consider the textures for $\mathbf m_u$ and $\mathbf m_d$ 
at $M_{GUT}$, the scale where they are presumably to be defined 
in view of a flavour theory.
The textures of eq. (\ref{ourtext}) 
are defined at $M_{GUT}=2\times 10^{16}$ GeV in the supersymmetric grand unified context 
and have to be run down to $m_Z$ to be compared to the experimental data, 
as we have done in the previous section.
The third family Yukawa couplings dominate in the renormalization group equation
and of course they affect the textures to some extent.

We have studied this effect that can be summarized as follows:\\
1) for small $\tan \beta$, in which case the top quark coupling dominates,
these changes are such to affect the textures and to slightly worsen the quality
of the fit, in particular for $|V_{cb}|$ which tends to be quite small;\\
2) for larger values of $\tan \beta$ the contribution due to the running of bottom and 
$\tau$ lepton couplings - assumed to be equal at $M_{GUT}$ -
corrects the effect due to the top coupling enough to allow a good fit to the data.
Therefore, the model prefers values of $\tan \beta$ above $15$.

\begin{figure}[!t]
\vskip .4cm
\footnotesize 
\begin{tabular}{c||c|c|c|c}
& & & & \\
$\tan \beta =5$ & $y_t$  & $y_b$ & $\mathbf T_u$ &  $\mathbf T_d^{1)}$ \\ & & & & \\ \hline \hline & & & & \\
$M_{GUT}$       & $0.61$ & $0.028$ & 
$\left( \matrix{  -3.70 \lambda^8 & 1.84 \lambda^6 & 0 \cr
              1.84\lambda^6 & 0 & 1.03 \lambda^2 \cr
              0 & 1.03 \lambda^2 & 1         } \right)$ & 
$\left( \matrix{ 0 & 0.33 \lambda^3 & 0 \cr
             0.33 \lambda^3 & i ~0.34 \lambda^2 & 0  \cr
             0  & 0.34 \lambda^2 & 1         } \right)$ \\ & & & & \\ \hline & & & & \\ 
$m_Z$           &  $1.00$   & $0.086$ & 
$\left( \matrix{  -5.49 \lambda^8 & 2.73 \lambda^6 & O( \lambda^8) \cr
              2.73 \lambda^6 & -0.51 \lambda^4  & 1.03 \lambda^2 \cr
              O( \lambda^8) & 1.03 \lambda^2 & 1         } \right)$ &  
$\left( \matrix{ O(\lambda^{15}) & 0.38 \lambda^3 & O(\lambda^7) \cr
             0.38 \lambda^3 & i ~0.39 \lambda^2 & O(\lambda^6)  \cr
             O(\lambda^9)  & 0.24 \lambda^2 & 1         } \right)$   \\ & & & & \\
\end{tabular}
\normalsize
\vskip 1. cm
\centerline{ \psfig{file=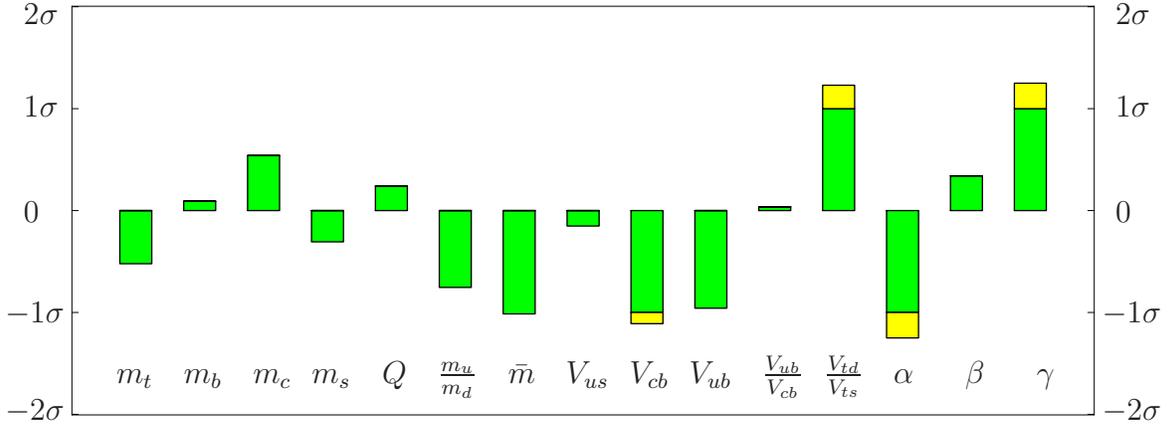,width=.85 \textwidth} 
\put(-370, 12){$m_t~~~m_b~~~m_c~~m_s~~~Q~~~\frac{m_u}{m_d}~~~\bar m~~~V_{us}~~V_{cb}~~V_{ub}
~~~\frac{V_{ub}}{V_{cb}}~~\frac{V_{td}}{V_{ts}}~~~\alpha~~~~~\beta~~~~~\gamma$}
\put(-412, -2){$-2\sigma$}\put(3, -2){$-2\sigma$}
\put(-412, 35){$-1\sigma$}\put(3, 35){$-1\sigma$}
\put(-405, 73){$0$}\put(8, 73){$0$}
\put(-405, 113){$1\sigma$}\put(8, 113){$1\sigma$}
\put(-405, 148){$2\sigma$}\put(8, 148){$2\sigma$}}
\caption{Predictions of the textures above with $\tan\beta=5$ faced to experimental data. 
For quark masses we use the PDG intervals \cite{PDGs} while for CKM we adopt the fit of \cite{ckmf}. 
The ellipse parameter $Q$ is consistent within 1$\sigma$ with the CPT result \cite{Leut}.}
\vskip 1. cm
\label{figfit5}\end{figure}

\begin{figure}[!t]
\vskip .4cm
\footnotesize             
\begin{tabular}{c||c|c|c|c}
& & & & \\
$\tan \beta =35$ & $y_t$  & $y_b$ & $\mathbf T_u$ &  $\mathbf T_d^{1)}$ \\ & & & & \\ \hline \hline & & & & \\
$M_{GUT}$       & $0.61$ & $0.233$ & 
$\left( \matrix{  -3.60 \lambda^8 & 1.77 \lambda^6 & 0 \cr
              1.77\lambda^6 & 0 & 1.01 \lambda^2 \cr
              0 & 1.01 \lambda^2 & 1         } \right)$ & 
$\left( \matrix{ 0 & 0.32 \lambda^3 & 0 \cr
             0.32 \lambda^3 & i ~0.33 \lambda^2 & 0  \cr
             0  & 0.33 \lambda^2 & 1         } \right)$ \\ & & & & \\ \hline & & & & \\ 
$m_Z$           &  $0.99$   & $0.589$ & 
$\left( \matrix{  -5.47 \lambda^8 & 2.69 \lambda^6 & O( \lambda^8) \cr
              2.69 \lambda^6 & -0.50 \lambda^4  & 1.01 \lambda^2 \cr
              O( \lambda^8) & 1.01 \lambda^2 & 1         } \right)$ &  
$\left( \matrix{ O(\lambda^{15}) & 0.40 \lambda^3 & O(\lambda^7) \cr
             0.40 \lambda^3 & i ~0.41 \lambda^2 & O(\lambda^6)  \cr
             O(\lambda^9)  & 0.24 \lambda^2 & 1         } \right)$   \\ & & & & \\
\end{tabular}
\normalsize
\vskip 1. cm
\centerline{ \psfig{file=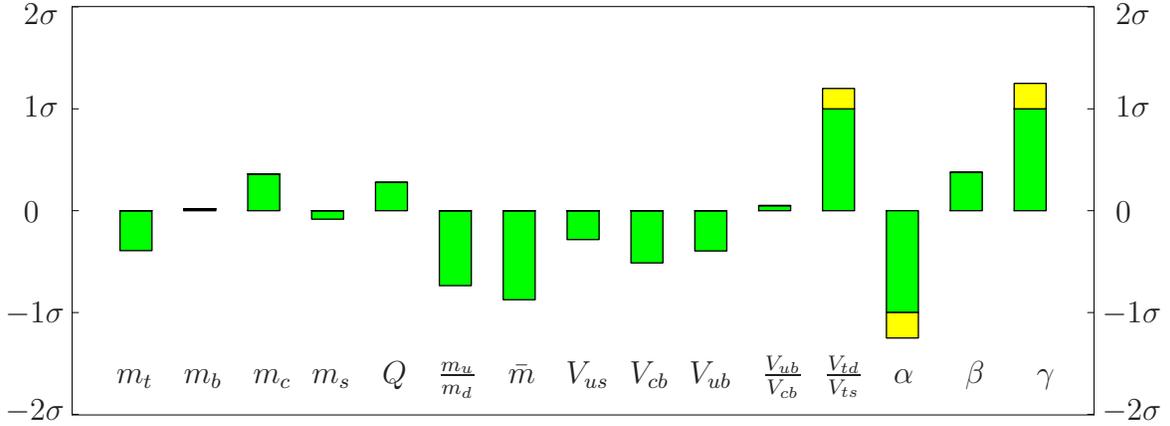,width=.85 \textwidth} 
\put(-370, 12){$m_t~~~m_b~~~m_c~~m_s~~~Q~~~\frac{m_u}{m_d}~~~\bar m~~~V_{us}~~V_{cb}~~V_{ub}
~~~\frac{V_{ub}}{V_{cb}}~~\frac{V_{td}}{V_{ts}}~~~\alpha~~~~~\beta~~~~~\gamma$}
\put(-412, -2){$-2\sigma$}\put(3, -2){$-2\sigma$}
\put(-412, 35){$-1\sigma$}\put(3, 35){$-1\sigma$}
\put(-405, 73){$0$}\put(8, 73){$0$}
\put(-405, 113){$1\sigma$}\put(8, 113){$1\sigma$}
\put(-405, 148){$2\sigma$}\put(8, 148){$2\sigma$}}
\caption{Predictions of the textures above with $\tan\beta=35$ faced to experimental data. 
For quark masses we use the
PDG intervals \cite{PDGs} while for CKM we adopt the fit of \cite{ckmf}. 
The ellipse parameter $Q$ is consistent within 1$\sigma$ with the CPT result \cite{Leut}.}
\vskip 1. cm
\label{figfit35}\end{figure}

Using the 1-loop renormalization group equations, we have carried out a quantitative study of the quark textures 
$\mathbf T_u$ and $\mathbf T_d^{1)}$ for the two representative cases $\tan\beta=5$ and $\tan\beta=35$. 
In both cases, we display the textures at $M_{GUT}$, where they are naturally defined, 
and show how they are modified at low energy, where quark masses and the CKM matrix elements are measured.   
As can be seen from figs. \ref{figfit5} and \ref{figfit35}, experimental data are well reproduced.
Since the angles of the unitarity triangle are mildly 
affected by the running (independently of the value of $\tan \beta$), 
the typical signature of our proposed textures, $\alpha\approx 90^\circ$, persists.

\section{Extension to the Lepton Sector}

The picture can now be completed by considering also the lepton sector.
Embedding $\mathbf T_d$ of eq. (\ref{ourtext}) in an $SU(5)$ grand unified theory, 
one realizes that the GJ relations 
$m_b=m_\tau$, $3 m_s=m_\mu$ and $m_d=3m_e$ \cite{GJ},
are achieved with
$(g,\hat g)_d \rightarrow (g, \hat g)_e$ while $(j,\hat j)_d \rightarrow (-3j,-3\hat j)_e$,
corresponding respectively to a $\bar{\underline{5}}_H$ and a $\bar{\underline{45}}_H$
(the latter could result from the insertion of a $\underline{24}_H$ or $\underline{75}_H$).
With the GJ prescription, the fit of the quark mass matrices of Section 5 leads to 
charged lepton masses that miss the experimental value by $O(10\%)$.

More generally, the parameters in $\mathbf T_d$ and $\mathbf T_e$ would get contributions
from higher dimension operators that would slightly modify the GJ relations \cite{BHRRU2}.
Let us decompose the values of the
coefficients found in fig. \ref{figfit35} for $\tan \beta =35$ 
in sums of $\bar{\underline{5}}_H$ and $\bar{\underline{45}}_H$ contributions:
$g=g_5+g_{45}$, $j=j_5+j_{45}$, $1=u_5+u_{45}$.
Hence, $\mathbf T_e^{1)T}$ has the same form of $\mathbf T_d^{1)}$ but with coefficients
$g_5-3g_{45}$, $j_5-3j_{45}$, $u_5-3u_{45}$, respectively. 
From the fit of the charged lepton masses run down at the scale $m_Z$, 
it turns out that $g_{45}/g_5= -3.8\%$, $j_{5}/j_{45}=-15.7\% $, $u_{45}/u_5=-4.5\%$.
However, it is well known that
$b-\tau$ unification \cite{btauoriginal} is a very model dependent issue:
there are significant low energy threshold corrections that
restrict the supersymmetric parameter space 
and strongly depend on $\tan \beta$ \cite{drastic}; also,
possible high energy threshold corrections could be significant, 
like e.g. the seesaw or GUT ones \cite{corrss}.  
Therefore, a detailed analysis of the extension of our textures to the charged lepton sector 
would only be justified in the context of a well defined supersymmetric model 
and a neutrino mass generation mechanism.  

Supported by the fact that the charged lepton textures turn out to be 
quite close to the simple GJ extension, let us analyse the consequences 
of such prescription for mixings and CPV in the lepton sector.
We write the two mass textures 
\bea
\tilde \mathbf T_d^{1)}&=&\mathbf T_e^{1)~T}=\left( \matrix{ 0 & g\lambda^3 & 0 \cr
             g\lambda^3 & -3ij\lambda^2 & 0   \cr
            0 &   -3j\lambda^2& 1         } \right) \label{ch1} \\
\tilde \mathbf T_d^{2)}&=&\mathbf T_e^{2)~T}= \left( \matrix{ 0 & \hat g \lambda^3 & 0 \cr
              \frac{\hat g}{\sqrt{2}} \lambda^3 & -3i\hat j \lambda^2 & i \cr
                i\frac{\hat g}{\sqrt{2}} \lambda^3 & 0 & 1         } \right)~ \frac{1}{\sqrt{2}}~,
                \label{ch2}
\eea
which differ by the magnitude of $\theta^{e_L}_{23}$,
while the other non-vanishing mixings are small and simply related to the left-mixings of down-quarks:
\bea
1) ~~s^{e_L,d_R}_{23} \simeq 0~~~~,~~~~~~~~~~~~
2)~~s^{e_L,d_R}_{23}\simeq \frac{1}{\sqrt{2}}~~~~,~~~~~~~~~~\\
\frac{m_\mu}{m_\tau}\simeq s^{e_R}_{23}=3 s^{d_L}_{23}~~~~~~~,~~~~~~~~~~~~~~
\sqrt{\frac{m_e}{m_\mu}}\simeq s^{e_L,e_R}_{12}=\frac{s^{d_R,d_L}_{12}}{3} \sim \frac{\theta_C}{3} ~~.
\eea
It is also interesting to study possible connections between CPV in quark and lepton sectors.
Due to the $SU(5)$ relation $\Phi_{e_L}= -\tilde \Phi_{d_R}$, where
the latter are the phases associated to $\tilde \mathbf T_d$,
quark phases explicitly appear in the expression for $U_{MNS}$, eq. (\ref{2U}).

In case 1), one can read from eq. (\ref{ratios}) that $\tilde \phi_{12}^{d_R}=\pi/2$,   
hence $\phi_{12}^{e_L}=-\pi/2=-\phi_{12}^q$. 
Due to $\theta_{23}^{d_R, e_L}\simeq 0$, 
one can get rid of $\phi_{23}^{e_L}$ by proper phase redefinitions of fields.
The only source of flavour violation in the charged lepton sector, 
$\theta^{e_L}_{12} \sim 4^\circ$, is too small to account for the solar angle.
Hence, not only the atmospheric, but also the solar angle 
has to come from $m_\nu^{eff}$ of eq. (\ref{lagr}).  

Case 2) is more interesting, because $\theta_{23}^{e_L}$ is maximal and so is $\phi^{e_L}_{23}$.
Indeed, eqs. (\ref{ratios}) and (\ref{sd23}) imply $\tilde \phi_{12,23}^{d_R}=\pi/2$, so that 
\beq
\phi_{12,23}^{\ell}+ \phi_{12,23}^{\nu}=\phi_{12,23}^{e_L}=-\frac{\pi}{2}=-\phi_{12}^q~~.
\eeq
If $\phi_{23}^\nu$ is very small or close to $\pi$, the maximal phase $\phi_{23}^\ell$
makes the atmospheric angle maximal independently of $\theta_{23}^\nu$ \cite{Isamax};
namely, $\theta_{atm}$ is maximal if the maximal CPV between charged leptons is not compensated
by a large CPV between the corresponding neutrinos.
Since $\theta^{e_L}_{12} \sim 4^\circ$,
the solar mixing angle must mainly come from $m_\nu^{eff}$.
If $\theta^\nu_{13}\simeq 0$, one has the prediction
$|U_{e3}| \simeq s^{e_L}_{12}/\sqrt{2} \approx 0.05$ and 
the relation
\beq
\theta_{12}^{MNS}=\theta_{12}^{\nu} + |U_{e3}| \cos \delta^{MNS}~~.
\eeq
The value ~$\theta^\nu_{12}=\arctan( 1/\sqrt{2})$, typical for tribimixing models \cite{tribi}, 
is particularly interesting because,
in this case, the experimental value of the solar angle 
forces $\delta^{MNS}$ to be maximal (see fig. 3 of ref. \cite{Isamax}).


\section{Conclusions}

We have derived two sets of fermion mass matrices adopting the following guidelines:\\
- the least number of parameters with the highest number of vanishing elements;\\
- phases of the form $n \pi/2$.\\
The quark mass matrices with only four parameters corresponding to the four mass ratios, 
eqs. (\ref{1text}) and (\ref{2text}),  
give a reasonable fit but for $m_u$, which turns out too large as compared to the 
chiral perturbation theory value \cite{Leut}.
This requires the introduction of one more small parameter $O(m_u)$ which shifts this mass down,
as in eq. (\ref{ourtext}).
The fit shows that some matrix elements are likely to be exactly equal 
or just related by factors of $\sqrt{2}$, 
favouring the interpretation of the textures in terms of a non-abelian flavour theory.

A crucial prediction of these textures is $\alpha \approx \pi/2$.
This is a consequence of our justified assumption $\theta_{13}^{d_L, u_L}=0$.
Indeed, in such a case it turns out (fig. \ref{fig4}) 
that the best choice to account for the observed CPV in the quark sector
is the presence of only one maximal phase, to be identified with the non-removable phase $\phi_{12}^q$.

The present direct measure suggests
$\alpha={99^\circ} ^{+12^\circ} _{-9^\circ}$ at $1\sigma$ \cite{alfadirect},
with a sensitivity already comparable to that of the recent fits of the CKM unitary matrix,
$\alpha={98.1^\circ} ^{+6.3^\circ} _{-7^\circ}$ at $1\sigma$ \cite{ckmf}.
Interestingly enough, B physics experiments already in progress will 
significantly improve the sensitivity to $\alpha$ in a few years.
Such experiments will provide a crucial test of these textures.
If in future 
$\alpha$ will stay close to its present central value but its sensitivity will allow to exclude $\alpha=\pi/2$,   
a suitable modification makes the textures still viable.
This is achieved for instance by filling the 31-element of $\mathbf T_d^{1)}$ 
with an additional purely imaginary small parameter, of $O(\lambda^5)$.
In the case that $\alpha$ will turn out to be considerably larger than its present central value, 
the model requires more than just a small correction.

A large value of $\tan \beta$ is slightly preferred
because in the running from $M_{GUT}$ down to $m_Z$ 
the effect of the contribution of the top coupling deteriorates the fit of $|V_{cb}|$, 
which can be corrected by the bottom and $\tau$ contributions to the running.
We have provided explicit numerical examples of our proposed textures at $M_{GUT}$
for two representative values of $\tan \beta$, figs. \ref{figfit5} and \ref{figfit35}.

Leptons have been incorporated in the analysis in the framework of an $SU(5)$ GUT
by a generalisation of the GJ mechanism,
via the insertion of an $SU(5)$ breaking v.e.v. in the element of $\mathbf T_d$ 
which bears the maximal phase inducing $\alpha\approx \pi/2$.
The quark and lepton CPVs would then have a common origin, 
further related to $SU(5)$ breaking.
The known phase $\delta$ in the CKM, which is known not to be suitable for baryogenesis is related
to a phase in the lepton mass matrices that could be operative in leptogenesis.

At a more theoretical side, the obvious question is whether these textures are naturally understood, namely 
whether they can be derived in a flavour theory. 
It turns out that there is a relatively simple model which can 
explain and even relate the parameters in the textures \cite{MasSa}.

\vskip .5cm


\section*{Acknowledgements}

I.M. thanks G. Altarelli for useful discussions.
We thank the Dep. of Physics of Rome1 and the CERN for hospitality during the completion of this work.
This project is partially supported by the RTN European Program MRTN-CT-2004-503369.

\vskip .5cm
\appendix 

\section*{Appendix: Input data}

We collect below intervals (central values in the middle) at 1 and 2$\sigma$ 
extracted from the CKM fit or ref. \cite{ckmf} at $m_Z$:

\begin{center}
\begin{tabular}{c|c}
$|V_{us}|$       &  $(.22422,.22523,.22625,.22723,.22823)$ \\ 
$|V_{cb}|$       & $(.04047,.04131,.04224,.04267,.04310)$  \\ 
$|V_{ub}| $      & $(.003708,.003804,.003899, .004001,.00415)$  \\ \hline 
$|V_{td}/V_{ts}|$              & $(.1717,.1884,.1982,.2095,.2243)$ \\
$|V_{ub}/V_{cb}|$              & $(.0874,.0899,.0923,.0955,.0994)$ \\
$R_u$            & $(.377,.387,.398,.409,.424)$   \\ \hline
$\gamma$         & $(43.5,52.7,58.6,65.4,74.3)$  \\
$\beta$          & $(21.10,22.45,23.22,23.96,24.92)$  \\
$\alpha$         & $(82.4,91.1,98.1,104.4,114.9)$  \\ 
\end{tabular}
\end{center}


Using the PDG \cite{PDGs} values and the multiplicative factors to go to $m_Z$ collected below, 
\begin{center}
\begin{tabular}{c|c|c|c}
          &    PDG at 1$\sigma$ \cite{PDGs}    & CPT \cite{Leut} & factor to go to $m_Z$ \cite{Raby}\\
\hline
$m_t(pole)$   & $172.7\pm 2.9$ GeV  & & \\
$m_t(m_t)$    & $163.5\pm 2.8$ GeV     & & $1.06$ \\
$m_b(m_b)$    & $4.25\pm 0.15$ GeV   & & $0.69$  \\
$m_c(m_c)$    & $1.3\pm0.10$ GeV     & & $0.56$  \\
$m_s(2 GeV)$  & $105\pm 25$ MeV      & & $0.65$  \\ 
$m_d(2 GeV)$ & $4-8$ MeV             & & $0.65$  \\
$m_u(2 GeV)$ & $1.5-4$ MeV           & & $0.65$  \\
$\bar m(2 GeV)$ & $3-5.5$ MeV        & & $0.65$  \\
$Q=\frac{m_s/m_d}{\sqrt{1-(m_u/m_d)^2}}$ & $21-25$ & $22.7\pm0.8$ & \\ 
$m_u/m_d$    & $0.3-0.7$        &  $0.553\pm0.043$   & \\  
$m_s/m_d$    & $17-22$  & $18.9\pm0.8$ & \\
\end{tabular}
\end{center}


\noindent we obtain the following intervals (central values in the middle) 
for mass ratios at $m_Z$ :
\bea
\frac{m_c}{m_t}= (3.81,4.20,4.60)\times 10^{-3} ~~~~,~~~
\frac{m_s}{m_b}= (1.71,2.33,2.99)\times 10^{-2} ~~~~~,~~~~\nn\\ \\
\frac{m_c}{m_s}= (7.95,10.67,15.08)~~,~~~ 
\frac{m_d}{m_s}=(.046,.051,.059)~~,~~~  
\frac{m_u}{m_c} =(0.91,2.4,5.2)\times 10^{-3}~.\nn 
\eea

\vskip 1cm



\end{document}